\documentclass[aps,superscriptaddress,nofootinbib,onecolumn,a4paper,11pt]{revtex4-2}
\usepackage{amsfonts,amsmath,amssymb}
\usepackage{graphicx}
\usepackage{natbib}
\usepackage{bm}
\usepackage{hyperref}
\usepackage{dcolumn}
\usepackage{graphicx}
\usepackage{color}
\usepackage{xfrac}
\usepackage{slashed}
\usepackage[usenames,dvipsnames]{xcolor}
\usepackage{soul}
\usepackage{verbatim}
\usepackage{ulem}
\usepackage{float}

\def\sthk{ \bar{\mbox{s}}_{\theta,k}}
\def\pthk{ \bar{\mbox{p}}_{\theta,k}}

\def\trmin{{\rm tr}}

\renewcommand{\thefootnote}{\arabic{footnote}}
\begin{document}

\title{Topological susceptibility and axion properties in the presence of a strong magnetic field within the three-flavor NJL model}
\author{J.P.~Carlomagno}
\email{carlomagno@fisica.unlp.edu.ar}
\affiliation{Instituto de F\'isica La Plata, CONICET $-$ Departamento de F\'isica, Facultad de Ciencias Exactas,
Universidad Nacional de La PLata, C.C. 67, (1900) La Plata, Argentina}
\affiliation{CONICET, Rivadavia 1917, (1033) Buenos Aires, Argentina}
\author{D.~G\'omez~Dumm}
\affiliation{Instituto de F\'isica La Plata, CONICET $-$ Departamento de F\'isica, Facultad de Ciencias Exactas,
Universidad Nacional de La PLata, C.C. 67, (1900) La Plata, Argentina}
\affiliation{CONICET, Rivadavia 1917, (1033) Buenos Aires, Argentina}
\author{N.N.~Scoccola}
\affiliation{CONICET, Rivadavia 1917, (1033) Buenos Aires, Argentina}
\affiliation{Physics Department, Comisi\'{o}n Nacional de Energ\'{\i}a
At\'{o}mica, Avenida del Libertador 8250, 1429 Buenos Aires,
Argentina}

\begin{abstract}
We analyze the topological susceptibility and the axion properties in the
presence of an external uniform magnetic field, considering a three flavor
NJL model that includes strong CP violation through a 't Hooft-like flavor
mixing term. Both thermal and finite density effects are studied for
magnetic fields up to 1~GeV$^2$, and the corresponding phase transitions are
analyzed. To capture the inverse magnetic catalysis effect at finite
temperatures and densities, a magnetic field-dependent coupling constant is
considered. Our analytical and numerical results are compared with those
previously obtained from lattice QCD, chiral perturbation theory and other
effective models.
\end{abstract}


\maketitle \hfill

\section{Introduction}

As well known, Quantum Chromodynamics (QCD) contains gauge field
configurations that carry topological charge~\cite{Belavin:1975fg}. These
configurations interpolate between different vacua of the gluonic sector,
giving rise to the so-called ``$\theta$ vacuum" \cite{Callan:1976je}. At the
level of the QCD Lagrangian, this nontrivial nature of the QCD vacuum
implies the presence of a so-called $\theta$-term of the form
\begin{equation}
{\cal L}_\theta \ = \ \theta_0 \, \frac{g^2}{32\pi^2}\; G_{\mu\nu}
\tilde G^{\mu\nu}\ ,
\label{axionlag}
\end{equation}
where $g$ is the strong coupling constant, and $G_{\mu\nu}$, $\tilde
G_{\mu\nu}$ stand for the gluon field tensor and its dual. In fact, in the
context of the full Standard Model, the coefficient $\theta_0$ can be
modified through a chiral rotation. Considering the weak interaction sector,
by diagonalizing the Yukawa-generated quark mass matrix $M_q$ one has
\begin{equation}
\theta_0 \ \to \ \theta \ = \ \theta_0 + {\rm arg}\; {\rm det}\; M_q\ .
\end{equation}
The parameter $\theta$ (which can be taken to run from 0 to $2\pi$, due to
the periodicity of the action) can be in principle experimentally
determined. Presently, measurements of the neutron electric dipole moment
lead to the constraint $|\theta|\lesssim 10^{-11}$. Since a nonzero value of
$\theta$ would imply the existence of CP violation in the strong interaction
sector, the measurement of such an unexpectedly low upper bound for the
value of $|\theta|$ is known as the ``strong CP problem''.

As proposed almost 50 years ago by Peccei and Quinn, a possible solution for
this puzzle can be achieved by invoking the existence of an additional
global U(1) chiral symmetry~\cite{Peccei:1977hh,Peccei:2006as}. The
spontaneous breakdown of this symmetry leads to the presence of an
associated Goldstone boson, the axion, which implies an extra anomalous
contribution to the Lagrangian. Thus, the resulting effective potential gets
minimized for a nonzero vacuum expectation value of the axion field, leading
finally to a cancellation of the $\theta$ term and solving the strong CP
problem. In practice, the above described mechanism can be basically
implemented by promoting the parameter $\theta$ to a dynamical field $a$,
normalized by a dimensionful ``decay constant'' $f_a$. The effective
potential will be given by a periodic function of $a/f_a$, becoming
minimized for $\langle a\rangle = 0$.

After the original introduction of the axion as a possible solution of the
strong CP problem, its relevance has also been discussed in other contexts
(for a recent review see e.g.\ Ref.~\cite{DiLuzio:2020wdo}). For example, it
has been proposed a candidate for cold dark matter. In this sense, the
temperature dependence of the axion mass would play an important role in the
estimation of its cosmic
abundance~\cite{Preskill:1982cy,Abbott:1982af,Dine:1982ah}. Axions might
also play a key role in astrophysics, in particular for the anomalous
cooling of neutron stars (see e.g.\
Ref.~\cite{Raffelt:2006cw,Sedrakian:2015krq}). In this framework, the
influence of nonzero density and background magnetic fields on the axion
properties deserve particular interest.

The strength of the topological charge fluctuations in the QCD vacuum is
quantified by the topological susceptibility $\chi_t$, which is defined as a
second derivative of the QCD partition function with respect to $\theta$ and
is shown to be proportional to the axion mass $m_a$. Different approaches to
non-perturbative QCD have been used to determine this quantity. For example,
within the framework of Nambu-Jona-Lasinio (NJL) type models, estimates of
the topological susceptibility and its temperature dependence have been
studied in Refs.~\cite{Fukushima:2001hr, Costa:2008dp,Contrera:2009hk}. On
the other hand, lattice QCD (LQCD) results for $\chi_t$ at zero and finite
temperature can be found in
Refs.~\cite{Borsanyi:2016ksw,Petreczky:2016vrs,Taniguchi:2016tjc}.

More recently, considerable attention has been paid to the modifications of
$\chi_t$ and axion properties induced by the presence of strong magnetic
fields. Besides its importance within the above mentioned astrophysical
context, it has been realized that the effect of magnetic fields on the
topological structure of the QCD vacuum can be significant for the study of
heavy ion collisions. In fact, the existence of a chirality imbalance
induced by topology, in the presence of strong magnetic fields produced in a
noncentral heavy ion collision, can lead to the so-called chiral magnetic
effect, according to which positive and negative charges get separated along
the magnetic field direction~\cite{Fukushima:2008xe,Kharzeev:2024zzm}. As
discussed in Ref.~\cite{Asakawa:2010bu}, it is important to take into
account the effects of both the temperature and the magnetic field on QCD
vacuum fluctuations. These effects have been analyzed using Chiral
Perturbation Theory
(ChPT)~\cite{Adhikari:2021lbl,Adhikari:2021jff,Adhikari:2022vqs}, which
should be adequate for relatively low values of the magnetic field. In
addition, results have been obtained using two-flavor versions of the
local~\cite{Bandyopadhyay:2019pml} and nonlocal NJL
model~\cite{Ali:2020jsy}. One of the aims of the present work is to extend
those studies by considering a three-flavor version of the local NJL model.
It is worth pointing out that in the SU(3) chiral NJL model the coupling
that controls the flavor mixing effects related with the U(1)$_A$ anomaly
can be better determined, taking into account the phenomenological values of
meson masses. Moreover, three-flavor models allow for a more consistent
comparison with very recent results obtained using LQCD simulations with 2+1
flavors~\cite{Brandt:2024gso}. Another important purpose of our work is to
complement previous studies by analyzing nonzero density effects which, as
already mentioned, might become relevant when axions are considered in
astrophysical contexts.

This article is organized as follows. Following
Ref.~\cite{Chatterjee:2014csa}, in Sec.~\ref{model}.A we review the general
formalism corresponding to a three-flavor NJL model at finite temperature
and chemical potential in the presence of a constant magnetic field,
including the mentioned $\theta$ field. Then, in Sec.~\ref{model}.B, we
derive the expressions needed to obtain the quantities of interest related
with the topological susceptibility and the axion properties. Numerical
results for these quantities are discussed in Sec.~\ref{results}. Finally,
in Sec.~\ref{summary} we summarize our results and present our main
conclusions.

\section{Theoretical formalism}
\label{model}

\subsection{Effective NJL lagrangian and mean field equations at zero temperature}

We consider a three-flavor NJL Lagrangian that includes scalar and
pseudoscalar chiral quark couplings as well as a 't Hooft six-fermion
interaction term, in the presence of an external electromagnetic field
${\cal A}_\mu$. Moreover, we take into account the coupling to a field
$\theta(x) = a(x)/f_a$ of the form given by Eq.~(\ref{axionlag}). This can
be effectively done through the 't Hooft term (which accounts for the chiral
U(1)$_A$ anomaly) by performing a chiral rotation of quarks fields by an
angle $\theta$. In this way, the effective Euclidean action is given
by~\cite{Boomsma:2009eh,Chatterjee:2014csa}
\begin{equation}
S_E \ = \ \int d^4x  \left\{ \bar{\psi}\left( -i\ \rlap/\!D  + \hat{m} \right)\psi -
G\sum_{a=0}^{8}\left[ \left( \bar{\psi}\lambda_{a}\psi \right)^{2}
    +\left( \bar{\psi}\,i\gamma_{5}\lambda_{a}\psi \right)^{2} \right]
+K \left(e^{i \theta}\, d_{+}+ e^{-i \theta} \,d_{-}\right)\right\}\ ,
\label{lf}
\end{equation}
where $G$ and $K$ are coupling constants, $\psi
=\left(\psi_u,\psi_d,\psi_s\right)^{T}$ stands for a quark three-flavor
vector, $\hat{m}=\mathrm{diag}\left( m_{u},m_{d},m_{s} \right)$ is the
corresponding current quark mass matrix, and $d_{\pm}=\det\left[
\bar{\psi}\left(1\pm\gamma_{5} \right)\psi \right]$. In addition,
$\lambda_{a}$ denote the Gell-Mann matrices, with
$\lambda_{0}=\sqrt{2/3}\,I$, where $I$ is the unit matrix in the three
flavor space. The coupling of quarks to the electromagnetic field is
implemented through the covariant derivative $D_{\mu}=\partial_\mu - i \hat
Q {\cal A}_{\mu}$, where $\hat Q=\mathrm{diag}\left( Q_{u},Q_{d},Q_{s}
\right)$ represents the quark electric charge matrix, i.e.\ $Q_u/2 = -Q_d =
- Q_s = e/3$, $e$ being the proton electric charge. In the present work we
consider a static and constant magnetic field in the $3$-direction.

We proceed by bosonizing the action in terms of scalar $\sigma_a(x)$ and
pseudoscalar $\pi_a(x)$ fields, introducing also corresponding auxiliary
$\mbox{s}_a(x)$ and $\mbox{p}_a(x)$ fields. Following a standard procedure,
we start from the partition function
\begin{equation}
 Z=\int D\bar{\psi} D\psi \ e^{-S_E} \, .
\end{equation}
By introducing functional delta functions, the scalar
($\bar{\psi}\lambda_{a}\psi$) and pseudoscalar
($\bar{\psi}i\gamma_{5}\lambda_{a}\psi$) currents in $S_E$ can be replaced
by $\mbox{s}_a(x)$ and $\mbox{p}_a(x)$, and one can perform the functional
integration on the fermionic fields $\psi$ and $\bar{\psi}$.
Then, to carry out the integration over the auxiliary fields we use the
stationary phase approximation (SPA), keeping the functions
$\tilde{\mbox{s}}_a(x)$ and $\tilde{\mbox{p}}_a(x)$ that minimize the
integrand of the partition function. This yields a set of coupled equations
among the bosonic fields, from which one can take $\tilde{\mbox{s}}_a(x)$
and $\tilde{\mbox{p}}_a(x)$ to be implicit functions of $\sigma_a(x)$ and
$\pi_a(x)$. Finally, we consider the mean field (MF) approximation,
expanding the bosonized action in powers of field fluctuations around the
corresponding translationally invariant mean field values $\bar{\sigma}_a$
and $\bar{\pi}_a$. Thus, we write
$\sigma_a(x)=\bar{\sigma}_a+\delta\sigma_a(x)$ and
$\pi_a(x)=\bar{\pi}_a+\delta\pi_a(x)$, where, due to charge conservation,
only $\bar{\sigma}_a$, $\bar{\pi}_a$ with $a=0,3,8$ can be nonzero. For
convenience, we introduce the notation
$\bar{\sigma}=\mathrm{diag}(\bar{\sigma}_u,\bar{\sigma}_d,\bar{\sigma}_s)=\lambda_0
\bar{\sigma}_0 + \lambda_3\bar{\sigma}_3+\lambda_8\bar{\sigma}_8$ and
$\bar{\pi}=\mathrm{diag}(\bar{\pi}_u,\bar{\pi}_d,\bar{\pi}_s)=\lambda_0
\bar{\pi}_0 + \lambda_3\bar{\pi}_3+\lambda_8\bar{\pi}_8$.

At the mean field level, the Euclidean action per unit volume reads
\begin{eqnarray}
\dfrac{\bar{S}_{\! E}^{\; \mathrm{bos}}}{V^{(4)}} \
&=&  - \dfrac{N_c}{V^{(4)}} \sum_{f} \int d^4x \, d^4x' \ \trmin_D\,\ln \left(\mathcal{S}^f_{x,x'}\right)^{-1} -
\dfrac{1}{2}\sum_f  \Big[ \,\bar{\sigma}_f \ \bar{\mbox{s}}_f + \bar{\pi}_f \ \bar{\mbox{p}}_f +
G \left( \bar{\mbox{s}}_f^2 + \bar{\mbox{p}}_f^2 \right) \Big]
\nonumber \\
&& + \ \frac{K}{4} \Big[ \cos\theta \left( \bar{\mbox{s}}_u\bar{\mbox{s}}_d\bar{\mbox{s}}_s
- \bar{\mbox{s}}_u\bar{\mbox{p}}_d\bar{\mbox{p}}_s
- \bar{\mbox{p}}_u\bar{\mbox{s}}_d\bar{\mbox{p}}_s
- \bar{\mbox{p}}_u\bar{\mbox{p}}_d\bar{\mbox{s}}_s \right) \nonumber \\
& & \qquad\quad -
\sin\theta \left( \bar{\mbox{p}}_u\bar{\mbox{p}}_d\bar{\mbox{p}}_s
- \bar{\mbox{p}}_u\bar{\mbox{s}}_d\bar{\mbox{s}}_s
- \bar{\mbox{s}}_u\bar{\mbox{p}}_d\bar{\mbox{s}}_s
- \bar{\mbox{s}}_u\bar{\mbox{s}}_d\bar{\mbox{p}}_s \right)
\Big]\ ,
\label{mfaction}
\end{eqnarray}
where $\trmin_D$ stands for trace in Dirac space, while
\begin{equation}
\left(\mathcal{S}^f_{x,x'}\right)^{-1} \ = \ \delta(x-x')\left[ -i (
\slashed \partial - i Q_f \slashed {\cal A} ) + M_{sf} + i \gamma_5\,M_{pf} \right]
\end{equation}
is the inverse mean field quark propagator for each flavor. Here, we have
used the definitions $M_{sf} = m_f+\bar{\sigma}_f$ and $M_{pf} =
\bar{\pi}_f$. Moreover, in Eq.~(\ref{mfaction}) $\bar{\mbox{s}}_f$ are the
values of the auxiliary fields at the mean field level within the SPA
approximation, i.e.\ $\bar{\mbox{s}}_f = \tilde{\mbox{s}}_f(\bar \sigma_a)$.
They satisfy the conditions
\begin{eqnarray}
\bar{\sigma}_i + 2 G \, \bar{\mbox{s}}_i - \frac{K}{4}\,\sum_{jk}\,|\epsilon_{ijk}| \left[  \left(\bar{\mbox{s}}_j \, \bar{\mbox{s}}_k -\bar{\mbox{p}}_j \ \bar{\mbox{p}}_k\right) \cos\theta
 +  2 \, \bar{\mbox{s}}_j \ \bar{\mbox{p}}_k \ \sin\theta \right] &=& 0 \ ,\nonumber \\
\bar{\pi}_i + 2 G \, \bar{\mbox{p}}_i \, -\, \frac{K}{4} \,\sum_{jk}\, |\epsilon_{ijk}| \left[  \left(\bar{\mbox{s}}_j \, \bar{\mbox{s}}_k -\bar{\mbox{p}}_j \ \bar{\mbox{p}}_k\right) \sin\theta
 -  2 \, \bar{\mbox{s}}_j \ \bar{\mbox{p}}_k \ \cos\theta \right] &=& 0 \ ,
\label{mfcond}
\end{eqnarray}
where the values $1,2,3$ for indices $i$, $j$ and $k$ are equivalent to labels $f=u,d,s$.
From the conditions $ \delta \bar{S}_{\! E}^{\; \mathrm{bos}} / \delta
\bar{\sigma}_f =0$ and $ \delta \bar{S}_{\! E}^{\; \mathrm{bos}} / \delta
\bar{\pi}_f =0$ one can get now the ``gap equations''
\begin{equation}
 \bar{\mbox{s}}_f \, = \, 2\,\langle \bar q_f q_f\rangle\ , \qquad \qquad
 \bar{\mbox{p}}_f \, = \, 2\,\langle \bar q_f i\gamma_5 q_f\rangle\ ,
\end{equation}
where the quark-antiquark condensates are given by
\begin{equation}
 \langle \bar q_f q_f\rangle \, = \, - \dfrac{N_c}{V^{(4)}} \int d^4x \ \trmin_D
 \big[ \mathcal{S}^f_{x,x} \big] \ , \qquad \qquad
 \langle \bar q_f i\gamma_5 q_f\rangle \, = \, - \dfrac{N_c}{V^{(4)}} \int d^4x \ \trmin_D
 \big[ i \gamma_5 \, \mathcal{S}^f_{x,x} \big]\ .
\end{equation}

As is well known, the quark propagator can be written in different
ways. For convenience we use
\cite{Miransky:2015ava,Andersen:2014xxa}
\begin{equation}
\mathcal{S}^f_{x,x'} \ = \ e^{i\Phi_f(x,x')} \, \int \dfrac{d^4 p}{(2\pi)^4}\ e^{i p\, (x-x')}\, \tilde{\mathcal{S}}_p^f \, ,
\label{sfx}
\end{equation}
where $\Phi_f(x,x')$ is the so-called Schwinger phase, which depends on the
gauge choice, and vanishes for $x \rightarrow x'$. We express
$\tilde{\mathcal{S}}_p^f$ in the Landau level form
\begin{eqnarray}
\!\!\!\!\!\!\!\!\!\tilde{\mathcal{S}}_p^f & = & 2\ e^{-\vec p^{\,2}_\perp/B_f} \
\sum_{k=0}^\infty \ \frac{(-1)^k }{M_{sf}^2 + M_{pf}^2 +  p_\parallel^2 + 2 k B_f}\nonumber
\\&&\!\!\!\!\!\!\!\!\!\!\!\!\!
 \times \left\{ \left( M_{sf} - i \gamma_5\, M_{pf} - p_\parallel \cdot \gamma_\parallel\right)
\left[\, \Gamma^+\,  L_{k}\Big(\frac{2 \vec p^{\,2}_\perp}{B_f}\Big) -
\Gamma^-\, L_{k-1}\Big(\frac{2 \vec p^{\,2}_\perp}{B_f}\Big)\right] + 2\,  \vec{p}_\perp \cdot
\vec{\gamma}_\perp  \, L^1_{k-1}\Big(\frac{2 \vec p^{\,2}_\perp}{B_f}\Big)\right\}\ .
\label{sfp_ll}
\end{eqnarray}
Here, $L_k^\alpha(x)$ are generalized Laguerre polynomials, with the
convention $L_{-1}^\alpha(x)=0$. We have also introduced the definitions
$s_f = {\rm sign} (Q_f B)$ and $B_f=|Q_fB|$, together with $\Gamma^\pm = (1
\pm i s_f \gamma_1 \gamma_2)/2$. ``Perpendicular'' and ``parallel'' gamma
matrices have been collected into vectors $\gamma_\perp =
(\gamma_1,\gamma_2)$ and $\gamma_\parallel = (\gamma_3,\gamma_4)$, and, in
the same way, we have defined vectors $p_\perp = (p_1,p_2)$ and $p_\parallel
= (p_3,p_4)$. Note that in our convention $\{\gamma_\mu,\gamma_\nu\}=-2
\delta_{\mu\nu}$.

The resulting expressions for $\bar{\mbox{s}}_f$ and $\bar{\mbox{p}}_f$ are divergent and have to be properly
regularized. We use here the magnetic field independent regularization
(MFIR) scheme, in which one subtracts from the unregulated integral the $B
\to 0$ limit and then adds it in a regulated form. In this way, we obtain
\begin{eqnarray}
\bar{\mbox{s}}_f = - 2 N_c \, M_{sf} \, I_{1f}^B \ ,  \qquad \qquad
\bar{\mbox{p}}_f = - 2 N_c \, M_{pf} \, I_{1f}^B \ ,
\label{spmf}
\end{eqnarray}
where
\begin{eqnarray}
I_{1f}^B \ = \ I_{1f}^0 + I_{1f}^{\rm mag}\ .
\end{eqnarray}
For the quantity $I_{1f}^{0}$ we use here a 3D cutoff regularization. Thus, we
have
\begin{equation}
I^{0}_{1f} \ = \ \dfrac{1}{2 \pi^2} \left[ \Lambda \sqrt{\bar M_f^2 + \Lambda^2} +
\bar M_f^2 \ \ln\left( \dfrac{\bar M_f}{\Lambda+ \sqrt{\bar M_f^2 + \Lambda^2}}\right) \right]\ ,
\label{I1freg}
\end{equation}
where $\bar M_f = \sqrt{M_{sf}^2 + M_{pf}^2}\,$. On the other hand, the ``magnetic
piece'' $I_{1f}^{\rm mag}$ is found to be given by
\begin{equation}
I_{1f}^{\rm mag} \ = \ \dfrac{B_f}{2\pi^2} \left[ \ln \Gamma(x_f) -
\left(x_f - \dfrac{1}{2}\right) \ln x_f + x_f - \dfrac{\ln{2\pi}}{2} \right] \ ,
\label{i1}
\end{equation}
where $x_f=\bar M_f^2 /(2B_f)$.

The associated Euclidean regularized action per unit volume is given by
\begin{eqnarray}
\dfrac{\bar{S}_{\! E}^{\; \mathrm{bos}}}{V^{(4)}}\, &=& \, \sum_f\,
\omega_f^B
 - \dfrac{1}{2}\sum_i\, \left[ \bar{\sigma}_i \ \bar{\mbox{s}}_i +
\bar{\pi}_i \ \bar{\mbox{p}}_i + G \left( \bar{\mbox{s}}_i^2 +
\bar{\mbox{p}}_i^2 \right) \right]
\nonumber \\
&& +\; \frac{K}{24} \sum_{ijk}\, |\epsilon_{ijk}| \Big[
\cos\theta \left( \bar{\mbox{s}}_i\bar{\mbox{s}}_j\bar{\mbox{s}}_k
- 3\, \bar{\mbox{s}}_i\bar{\mbox{p}}_j\bar{\mbox{p}}_k \right)  -
\sin\theta \left( \bar{\mbox{p}}_i\bar{\mbox{p}}_j\bar{\mbox{p}}_k
- 3\, \bar{\mbox{p}}_i\bar{\mbox{s}}_j\bar{\mbox{s}}_k \right)
\Big]\ ,
\label{actionB}
\end{eqnarray}
where
\begin{eqnarray}
\omega_f^B \ = \ \omega_f^0 + \omega_f^{\rm mag}\ ,
\label{omegasuma}
\end{eqnarray}
with
\begin{eqnarray}
\omega_f^0 &=& - \frac{N_c}{8\pi^2} \left[\Lambda \sqrt{\bar M_f^2 + \Lambda^2} \left( \bar M_f^2 + 2 \Lambda^2 \right) +
\bar M_f^4 \ln\left( \frac{ \bar M_f}{\Lambda + \sqrt{\bar M_f^2 + \Lambda^2} }\right) \right]\ ,
\label{omegacero}\\
\omega_f^{\rm mag} &=& - \frac{N_c \, B_f^2}{2\pi^2} \left[ \zeta'(-1,x_f) - \frac{x_f^2 - x_f}{2} \ln x_f + \frac{x_f^2}{4} \right]\ .
\label{omegamag}
\end{eqnarray}
Here $\zeta'(-1,x_f)$ stands for the derivative of the Hurwitz zeta
function.

\hfill

Let us now consider the case of a system in equilibrium at a finite
temperature $T$ and quark chemical potential $\mu$. We follow here a similar
analysis as the one carried out in Ref.~\cite{Chatterjee:2014csa}. The
expressions for the mean field values $\bar s_f$ and $\bar p_f$ can be
written as in Eqs.~(\ref{spmf}), replacing the function $I_{1f}^B$ by
\begin{eqnarray}
I_{1f}^{B,T,\mu} \ = \ I_{1f}^B \, + \, I_{1f}^{{\rm mag},T,\mu}\ .
\end{eqnarray}
where
\begin{eqnarray}
I_{1f}^{{\rm mag},T,\mu}\, &=& \,\frac{B_f}{4 \pi^2} \sum_{k=0}^\infty \,\alpha_k
\int_{-\infty}^\infty dp \ \frac{1}{E_{kpf}}\, \sum_{s=\pm} \,
\frac{1}{ 1+ \exp[ (E_{kpf} + s \mu)/T]}\ ,
\end{eqnarray}
with $\alpha_k = 2 - \delta_{k0}$, $E_{kpf} = \sqrt{ p^2 +  2 k B_f + \bar M_f^2}\,$.

In the same way, the associated regularized thermodynamical
potential $\Omega$ can be expressed as in Eq.~(\ref{actionB}), replacing
$\omega^B_f$ by $\omega_f^{B,T,\mu}$,
where
\begin{eqnarray}
\omega_f^{B,T,\mu} \ = \ \omega_f^B \,+\, \omega_f^{{\rm mag},T,\mu}\ .
\end{eqnarray}
The $T=0$, $\mu=0$ piece $\omega_f^B$ is given by
Eqs.~(\ref{omegasuma}-\ref{omegamag}), while the finite $T,\mu$ piece reads
\begin{equation}
\omega_f^{{\rm mag},T,\mu}\ = \ - \frac{N_c \, T}{4 \pi^2} \sum_f B_f
\,\sum_{k=0}^\infty \, \alpha_k \int_{-\infty}^\infty dp \, \sum_{s=\pm} \,
\ln \Big\{ 1+ \exp[ -(E_{kpf} + s \mu)/T] \Big\}\ .
\end{equation}

\subsection{Mean field topological susceptibility, axion mass and axion self-coupling}

As stated, the topological susceptibility $\chi_t$ is useful to analyze how the
manifestations of the chiral anomaly are affected by the temperature and the
external magnetic field. This quantity is given by
\begin{equation}
\chi_t\ = \ \int\ d^4x\, \langle 0|\,T\,Q(x)\,Q(0)|0\rangle\ ,
\end{equation}
where $Q(x)$ is the topological charge
\begin{equation}
Q(x) \ = \ \frac{g^2}{32\pi^2}\; G_{\mu\nu}(x)\,
\tilde G^{\mu\nu}(x)\ .
\end{equation}
In the framework of the above introduced effective NJL model, $\chi_t$ can
be simply calculated by taking the second derivative of the effective action
with respect to $\theta$, viz.
\begin{equation}
\chi_t \ = \ \frac{d^2 \Omega}{d \theta^2}\Big|_{\theta =0}\ .
\end{equation}

The evaluation of the first derivative of $\Omega$ with respect to $\theta$
can be obtained taking into account the SPA equations
\begin{equation}
\frac{\partial \Omega}{\partial\bar{\mbox{s}}_f}\Big|_{s_f=\tilde s_f} \ =
\ \frac{\partial \Omega}{\partial \bar{\mbox{p}}_f}\Big|_{s_f=\tilde s_f}\ =
\ 0
\end{equation}
together with the MF conditions
\begin{equation}
\frac{\partial \Omega}{\partial \sigma_f}\Big|_{\sigma_f=\bar \sigma_f}
\ =\ \frac{\partial \Omega}{\partial \pi_f}\Big|_{\pi_f=\bar \pi_f} \
= \ 0\ .
\end{equation}
Then, from Eq.~(\ref{mfaction}) one has
\begin{equation}
\frac{d \Omega}{d \theta} \ = \ \frac{\partial \Omega}{\partial \theta}\ =
\ - \frac{K}{24}\, \sum_{ijk}\, |\epsilon_{ijk}|\,
\Big[ \sin\theta \left( \bar{\mbox{s}}_i\bar{\mbox{s}}_j\bar{\mbox{s}}_k
- 3\,\bar{\mbox{s}}_i\bar{\mbox{p}}_j\bar{\mbox{p}}_k\right) +
\cos\theta \left( \bar{\mbox{p}}_i\bar{\mbox{p}}_j\bar{\mbox{p}}_k
- 3\,\bar{\mbox{p}}_i\bar{\mbox{s}}_j\bar{\mbox{s}}_k\right)\Big]\ .
\end{equation}
Using Eqs.~(\ref{mfcond}), and noting that $\bar{\mbox{s}}_f\, M_{pf}
= \bar{\mbox{p}}_f\,M_{sf}$ (see Eqs.~(\ref{spmf})), one gets
\begin{equation}
\frac{d \Omega}{d \theta} \ = \ -\,\frac{1}{6}\sum_f \,m_f\,\bar{\mbox{p}}_f \ ,
\label{der1}
\end{equation}
where $m_f$ are the current quark masses. Moreover, it can be shown that the
terms in the above sum over flavors are equal to each other. Indeed, from
Eqs.~(\ref{mfcond}) one has
\begin{eqnarray}
m_i\, \bar{\mbox{p}}_i & \, = \, & \frac{K}{4}\,\sum_{jk}\,|\epsilon_{ijk}| \Big[
\,\bar{\mbox{p}}_i \, \bar{\mbox{p}}_j\, \bar{\mbox{p}}_k\,\cos\theta
\, + \, \bar{\mbox{s}}_i \, \bar{\mbox{s}}_j\, \bar{\mbox{s}}_k \sin\theta
\nonumber \\
& & \, - \left(\bar{\mbox{p}}_i \, \bar{\mbox{s}}_j\, \bar{\mbox{s}}_k +
2\,\bar{\mbox{s}}_i \, \bar{\mbox{p}}_j\, \bar{\mbox{s}}_k\right)\cos\theta
\, - \, \left(\bar{\mbox{s}}_i \, \bar{\mbox{p}}_j\, \bar{\mbox{p}}_k +
2\,\bar{\mbox{p}}_i \, \bar{\mbox{s}}_j\, \bar{\mbox{p}}_k\right)\sin\theta
\,\Big] \nonumber \\
& \, = \, & \frac{K}{12}\,\sum_{ljk}\,|\epsilon_{ljk}| \Big[
\,\bar{\mbox{p}}_l \, \left(\bar{\mbox{p}}_j\, \bar{\mbox{p}}_k
- 3 \bar{\mbox{s}}_j\, \bar{\mbox{s}}_k\right)\cos\theta \, + \,
\bar{\mbox{s}}_l \, \left(\bar{\mbox{s}}_j\, \bar{\mbox{s}}_k
- 3 \bar{\mbox{p}}_j\, \bar{\mbox{p}}_k\right)\sin\theta \, \Big]\ ,
\end{eqnarray}
where the last expression does not depend on $i$. To derive this relation we
have used the property
\begin{equation}
\sum_{jk}\,|\epsilon_{ijk}| \, (a_i\,b_j\, b_k + 2\, b_i\, a_j\, b_k) \ = \
\sum_{ljk}\,|\epsilon_{ljk}|\, a_l\,b_j\, b_k\ .
\end{equation}
Thus, we have
\begin{equation}
\frac{d \Omega}{d \theta} \ = \ -\,\frac{1}{2} \;m_f\,\bar{\mbox{p}}_f \ ,
\end{equation}
where $f$ can be either $u$, $d$ or $s$.

Notice that if any of the current quark masses (e.g.\ $m_u$) is taken to be
equal to zero, one immediately obtains $d \Omega/d \theta = 0$, i.e., the
Lagrangian in Eq.~(\ref{lf}) becomes independent of $\theta$. This is due to
the existence of an additional U(1) global symmetry. In this limit the value
of $\theta$ becomes unobservable, and, as is well known, the strong CP
problem vanishes.

{}From the above equations it is easy to see that, as required by the
Peccei-Quinn mechanism, the minimization condition $d\Omega/d\theta = 0$
leads to the mean field values $\bar\pi_f=0$, $\bar\theta = 0$. On the other
hand, at the mean field level, the second derivative of the action with
respect to $a = f_a\,\theta$ is nothing but the axion mass squared. Thus, the
topological susceptibility and the axion mass are simply related by
\begin{equation}
f_a^2\,m_a^2\ = \ \frac{d^2 \Omega}{d \theta^2}\Big|_{\theta =\bar\theta = 0}\ = \ \chi_t\ .
\end{equation}

To evaluate the second derivative in the above expression we need to
determine $d\bar{\mbox{p}}_f/d\theta$. We have
\begin{equation}
\frac{d \bar{\mbox{p}}_f}{d\theta} \ = \
\frac{\partial \bar{\mbox{p}}_f}{\partial\bar\sigma_f}\,
\frac{\partial \bar\sigma_f}{\partial\theta}\, + \,
\frac{\partial \bar{\mbox{p}}_f}{\partial\bar\pi_f}\,
\frac{\partial \bar\pi_f}{\partial\theta}\ .
\end{equation}
The expressions for $\partial \bar{\mbox{p}}_f/\partial\bar\sigma_f$ and
$\partial \bar{\mbox{p}}_f/\partial\bar\pi_f$ can be readily obtained from
Eqs.~(\ref{spmf}) and the subsequent expressions of the functions $I_{1f}$
for zero and nonzero temperature. The partial derivatives of $\bar\sigma_f$
and $\bar\pi_f$ with respect to $\theta$ can be calculated by solving the
coupled equations
\begin{eqnarray}
\frac{\partial \bar\sigma_i}{\partial\theta} \, +
\sum_{\phi = \bar\sigma,\bar\pi} \bigg[
2 G \,\frac{\partial \bar{\mbox{s}}_i}{\partial \phi_i}\,
\frac{\partial \phi_i}{\partial\theta} \, - \, \frac{K}{2} \,
\sum_{jk}\,|\epsilon_{ijk}|\,
\bigg(\frac{\partial \bar{\mbox{s}}_j}{\partial \phi_j}\, \bar{\mbox{s}}_{\theta,k}
- \frac{\partial \bar{\mbox{p}}_j}{\partial \phi_j}\,
\bar{\mbox{p}}_{\theta,k}\bigg)\frac{\partial \phi_j}{\partial\theta}\bigg]
& \, = \, & -\,\bar\pi_i\, -\, 2 G\, \bar{\mbox{p}}_i\ ,
\nonumber \\ [2mm]
\frac{\partial \bar\pi_i}{\partial\theta} \, +
\sum_{\phi = \bar\sigma,\bar\pi} \bigg[
2 G \,\frac{\partial \bar{\mbox{p}}_i}{\partial \phi_i}\,
\frac{\partial \phi_i}{\partial\theta} \, + \, \frac{K}{2} \,
\sum_{jk}\,|\epsilon_{ijk}|\,
\bigg(\frac{\partial \bar{\mbox{s}}_j}{\partial \phi_j}\, \bar{\mbox{p}}_{\theta,k}
+ \frac{\partial \bar{\mbox{p}}_j}{\partial \phi_j}\,
\bar{\mbox{s}}_{\theta,k}\bigg)\frac{\partial \phi_j}{\partial\theta}\bigg]
& \, = \, & \bar\sigma_i\, +\, 2 G\, \bar{\mbox{s}}_i\ ,
\end{eqnarray}
where we have introduced the shorthand notation
\begin{equation}
\sthk = \bar{\mbox{s}}_k\cos\theta + \bar{\mbox{p}}_k\sin\theta
\ ,\qquad\qquad
\pthk = \bar{\mbox{p}}_k\cos\theta - \bar{\mbox{s}}_k\sin\theta\ .
\end{equation}
In the limit $\theta = 0$ the equations get simplified, leading to the
result
\begin{equation}
\chi_t \ = \ \frac{d^2 \Omega}{d \theta^2}\Big|_{\theta = 0}\ = \
- \frac{1}{2}\; \bigg[\, \frac{2}{K\,\bar s_u\,\bar s_d\,\bar s_s}
\, + \,
\sum_{k} \, \frac{1}{m_k\,\bar s_k}\,\bigg]^{-1}\ .
\label{chit_exacta}
\end{equation}
We recall that $\bar s_f$ is equal to twice the scalar condensate $\langle \bar q_f
q_f\rangle$.

The expression in Eq.~(\ref{chit_exacta}) can be compared to previous
results obtained in the approximate chiral limit, where current masses are
taken to be relatively small. At the lowest order in $m_f^{-1}$ one has
\begin{equation}
\chi_t \ \simeq \ - \frac{1}{2}\;
\bigg(\,
\sum_{k} \, \frac{1}{m_k\,\bar s_k}\,\bigg)^{-1}\ .
\label{chit_app1}
\end{equation}
Moreover, assuming $\bar s_u\simeq \bar s_d\simeq \bar s_s$ one can
approximate
\begin{equation}
\chi_t \ \simeq
 \ - \; \frac{\displaystyle
\frac{1}{2}\,\sum_k \frac{\bar s_k}{m_k}}{\displaystyle \rule{0cm}{5.9mm}
\Big(\,\sum_{k} \, \frac{1}{m_k}\,\Big)^2}\ +
\ {\cal O}\Big(\Delta_{su}^2\,,\Delta_{du}\Delta_{su}\,,\,\Delta_{du}^2\Big)\ ,
\label{chit_app2}
\end{equation}
where $\Delta_{fu} = (\bar s_f-\bar s_u)/\bar s_u$. This in agreement with
the expressions found in Refs.~\cite{Mao:2009sy,Adhikari:2022vqs}
and~\cite{Kawaguchi:2020qvg} in the contexts of ChPT and a linear sigma
model, respectively. In the limit $\bar s_u = \bar s_d = \bar s_s$, the
above expressions for $\chi_t$ reduce to the lowest-order ChPT
Leutwyler-Smilga relation~\cite{Leutwyler:1992yt}
\begin{equation}
\chi_t \ \simeq \ - \; \frac{\bar s_f}{2}\,
\Big(\,\sum_{k} \, \frac{1}{m_k}\,\Big)^{-1}\ .
\label{chit_app3}
\end{equation}

Finally, it is also interesting to study the axion self-coupling arising
from the $\theta$-dependent effective potential. It is usual to focus on the
$\theta^4$ (i.e.~$(a/f_a)^4$) term in the effective action, defining the
coupling parameter $\lambda_a$ as
\begin{equation}
\lambda_a \ = \ \frac{1}{f_a^4}\;\frac{d^4 \Omega}{d \theta^4}\Big|_{\theta = 0}\ .
\end{equation}

\section{Results}
\label{results}

\subsection{Model parameters}

Before presenting the numerical results for the topological susceptibility,
we introduce the parameterization that has been chosen for the above
discussed three-flavor version of the NJL model. Following
Refs.~\cite{Rehberg:1995kh, Coppola:2024uvz}, we adopt the parameter set
$m_u = m_d = 5.5$~MeV, $m_s = 140.7$~MeV, $\Lambda = 602.3$~MeV, $G\Lambda^2
= 1.835$ and $K\Lambda^5 = 12.36$. This set has been determined in such a
way that for $B=T=0$ one obtains meson masses $m_\pi = 135$~MeV, $m_K =
497.7$~MeV and $m_{\eta'} = 957.8$~MeV, together with a pion decay constant
$f_\pi = 92.4$~MeV.

\begin{figure}[hbt]
\centering
\includegraphics[width=0.6\textwidth]{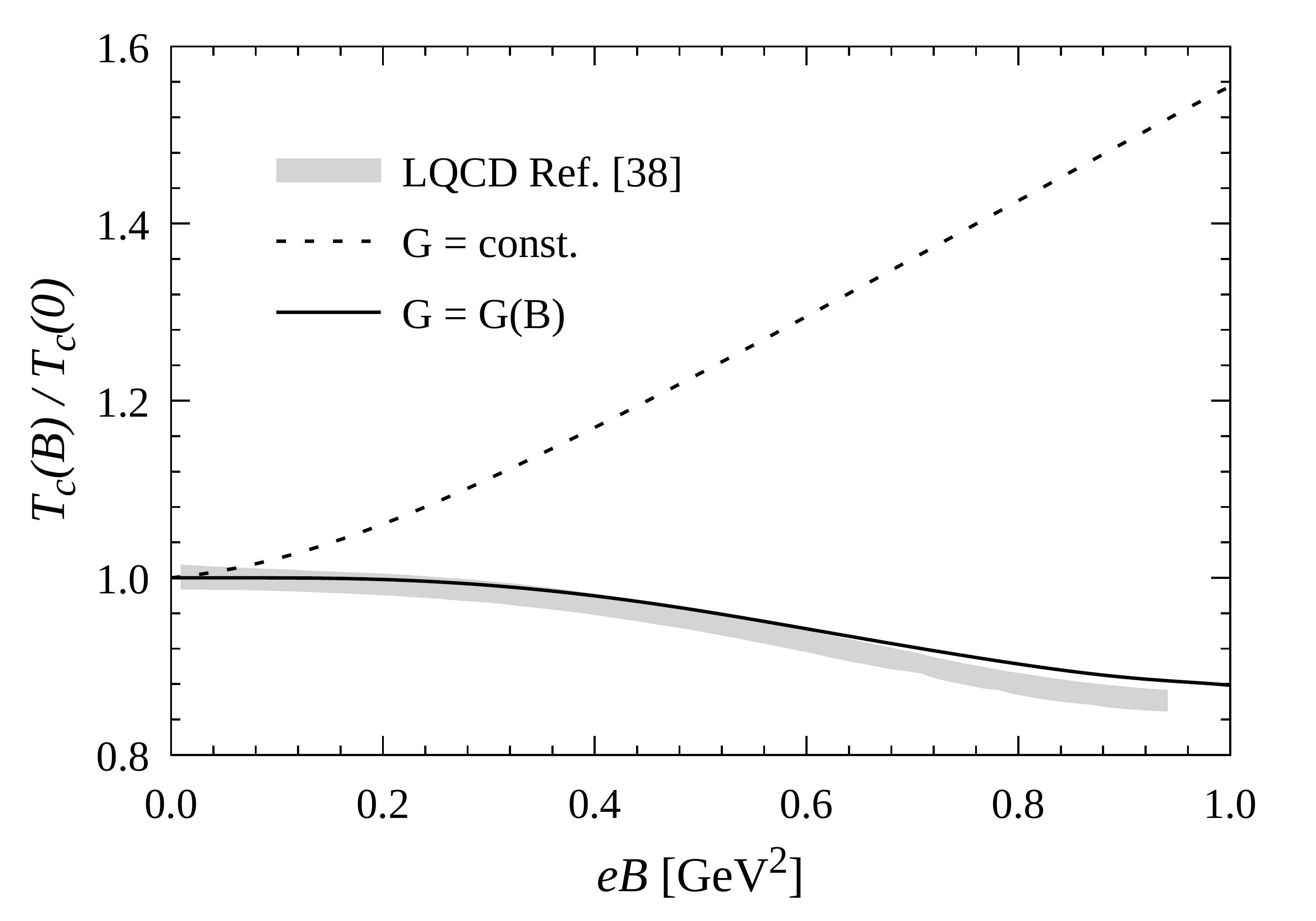}
\caption{Normalized values of the pseudocritical chiral restoration
temperatures as functions of the magnetic field, for constant and
$B$-dependent $G$~\cite{Ferreira:2014kpa}. Lattice QCD results from
Ref.~\cite{Bali:2011qj} are included for comparison.}
\label{fig1}
\end{figure}

It is well known that local NJL-like models fail to reproduce the inverse
magnetic catalysis (IMC) effect at finite temperature. To address this
issue, the possibility of allowing the coupling constant $G$ to
depend on the magnetic field has been
considered~\cite{Ferreira:2014kpa,Farias:2014eca,Farias:2016gmy}.
Taking into account the analysis carried out in
Ref.~\cite{Ferreira:2014kpa}, we consider a functional form
\begin{align}
G(B) \:=\: G \, \left[ \dfrac{ 1 + a \, (eB/\Lambda^2_{\rm QCD})^2 + b \, (eB/\Lambda^2_{\rm QCD})^3 }
                             { 1 + c \, (eB/\Lambda^2_{\rm QCD})^2 + d \, (eB/\Lambda^2_{\rm QCD})^4 } \right] \ ,
\label{gdeb}
\end{align}
where the parameters $a$, $b$, $c$ and $d$ are determined by fitting the $B$
dependence of the pseudocritical chiral transition temperatures to those
obtained through LQCD calculations~\cite{Bali:2011qj}. The values of the
parameters are~\cite{Ferreira:2014kpa} $a = 0.0108805$, $b = -1.0133 \times
10^{-4}$, $c = 0.02228$ and $d = 1.84558 \times 10^{-4}$, with $\Lambda_{\rm
QCD} = 300$~MeV. The effect on the pseudocritical chiral restoration
temperatures $T_c$ (at zero baryon chemical potential) is shown in
Fig.~\ref{fig1}, where we plot $T_c$ ---normalized to $T_c(B=0)$--- as a
function of the magnetic field, considering the case of a constant coupling
$G$ and the case in which one has a $B$-dependent coupling $G(B)$ as in
Eq.~(\ref{gdeb})~\cite{Ferreira:2014kpa}. For comparison, the $B$ dependence
of the normalized pseudocritical temperatures obtained from Lattice QCD
calculations is also shown (gray band in the figure)~\cite{Bali:2011qj}. The
normalization temperature in our model is found to be $T_c(B=0)= 173$~MeV,
somewhat larger than the critical value obtained from LQCD calculations,
$T_c(B=0)= 156$~MeV~\cite{HotQCD:2018pds}.

\subsection{Zero chemical potential}

\begin{figure}[hbt]
\centering
\includegraphics[width=0.55\textwidth]{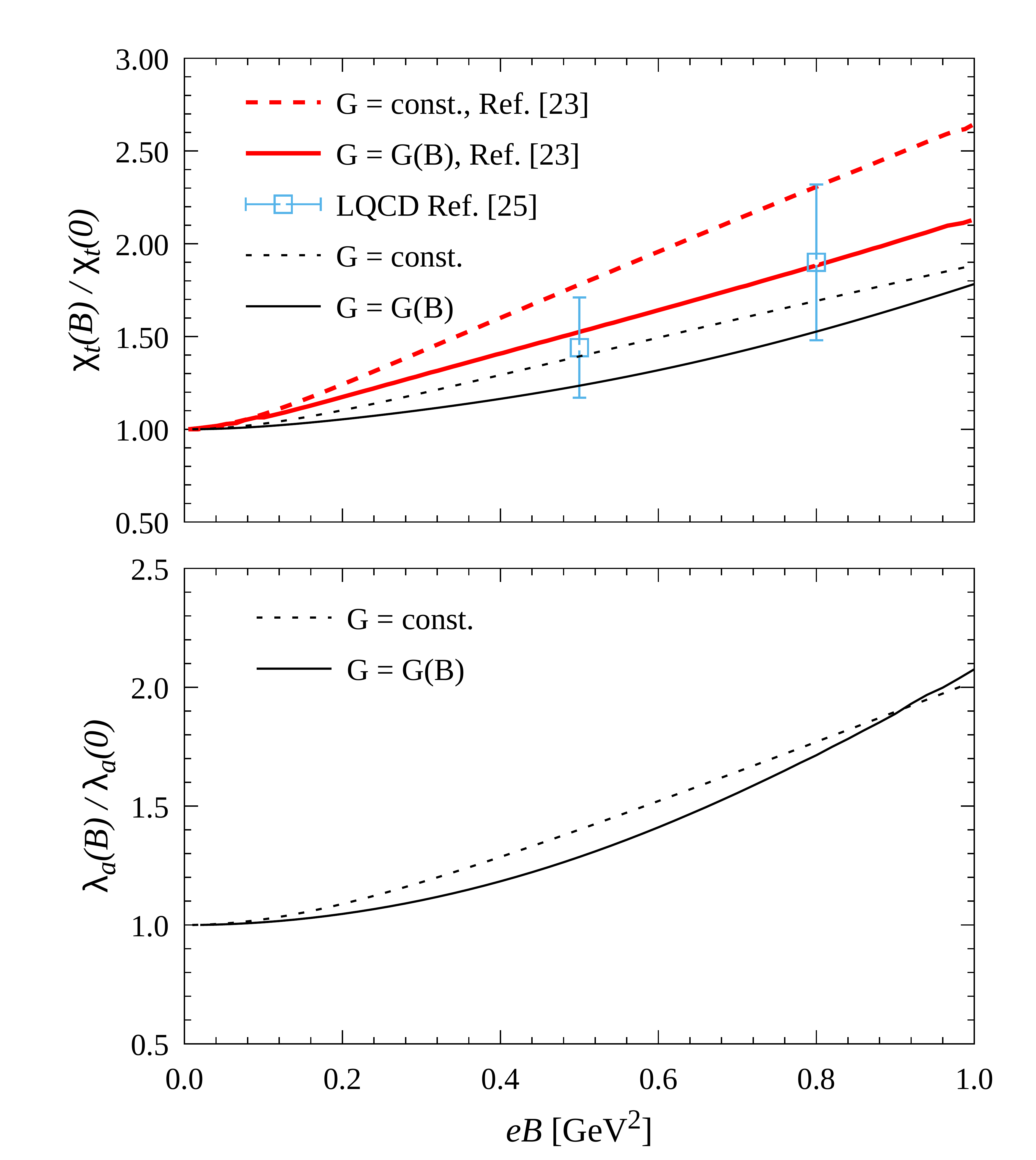}
\caption{Normalized values of $\chi_t$ and $\lambda_a$ as functions of $eB$
for $\mu=T=0$. The cases $G={\rm constant}$ and $G=G(B)$ are considered.
Results from a two-flavor NJL model~\cite{Bandyopadhyay:2019pml} and
LQCD~\cite{Brandt:2024gso} are shown for comparison.}
\label{fig2}
\end{figure}

Let us start by quoting our numerical results for both zero quark chemical
potential and zero temperature. In Fig.~\ref{fig2} we show the values of the
topological susceptibility $\chi_t$ (upper panel) and the axion
self-coupling parameter $\lambda_a$ (lower panel) as functions of the
magnetic field, normalized by the corresponding values at $B=0$, namely
$\chi_t(B=0) = 78$~MeV and $\lambda_a(B=0) = 0.85\times
10^{-5}$~GeV$^4/f_a^4$. Black dashed and solid lines correspond to $G = {\rm
constant}$ and $G=G(B)$, respectively. It can be seen that in both cases
$\chi_t$ and $\lambda_a$ show an enhancement with the magnetic field. Within
errors, our results for the topological susceptibility are shown to be in
agreement with those obtained from LQCD calculations (blue
squares)~\cite{Brandt:2024gso} for a temperature $T\simeq 110$~MeV (which is
well below the critical temperature, therefore the value of $\chi_t$ should
be rather close to the one at $T=0$). In addition, we include for comparison
the curves for $\chi_t$ corresponding to a two-flavor NJL model, taken from
Ref.~\cite{Bandyopadhyay:2019pml}, both for constant and $B$-dependent
couplings (red dashed and solid lines, respectively). The above mentioned
value for $\chi_t$ at $B=0$ obtained within our model can be compared with
the results obtained from ChPT~\cite{Gorghetto:2018ocs} and
LQCD~\cite{Borsanyi:2016ksw} analyses, which lead to $\chi_t(B=0)\simeq
75.5$~MeV. In the case of the axion self-coupling, from ChPT the estimation
$\lambda_a\simeq 1.12\times 10^{-5}$~GeV$^4/f_a^4$ is
obtained~\cite{GrillidiCortona:2015jxo}.

\begin{figure}[hbt]
\centering
\includegraphics[width=0.49\textwidth]{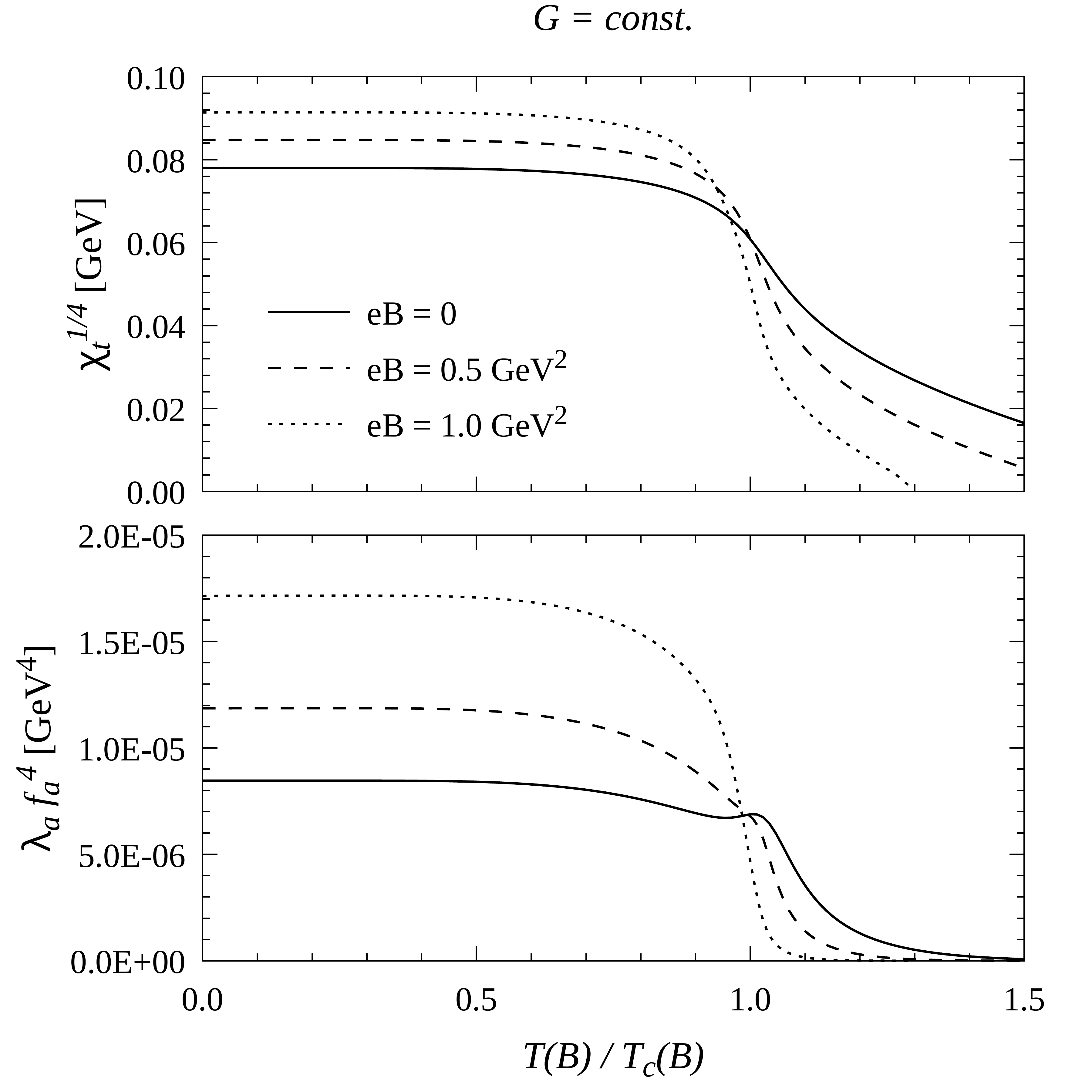}
\hspace*{-1.5cm}
\includegraphics[width=0.49\textwidth]{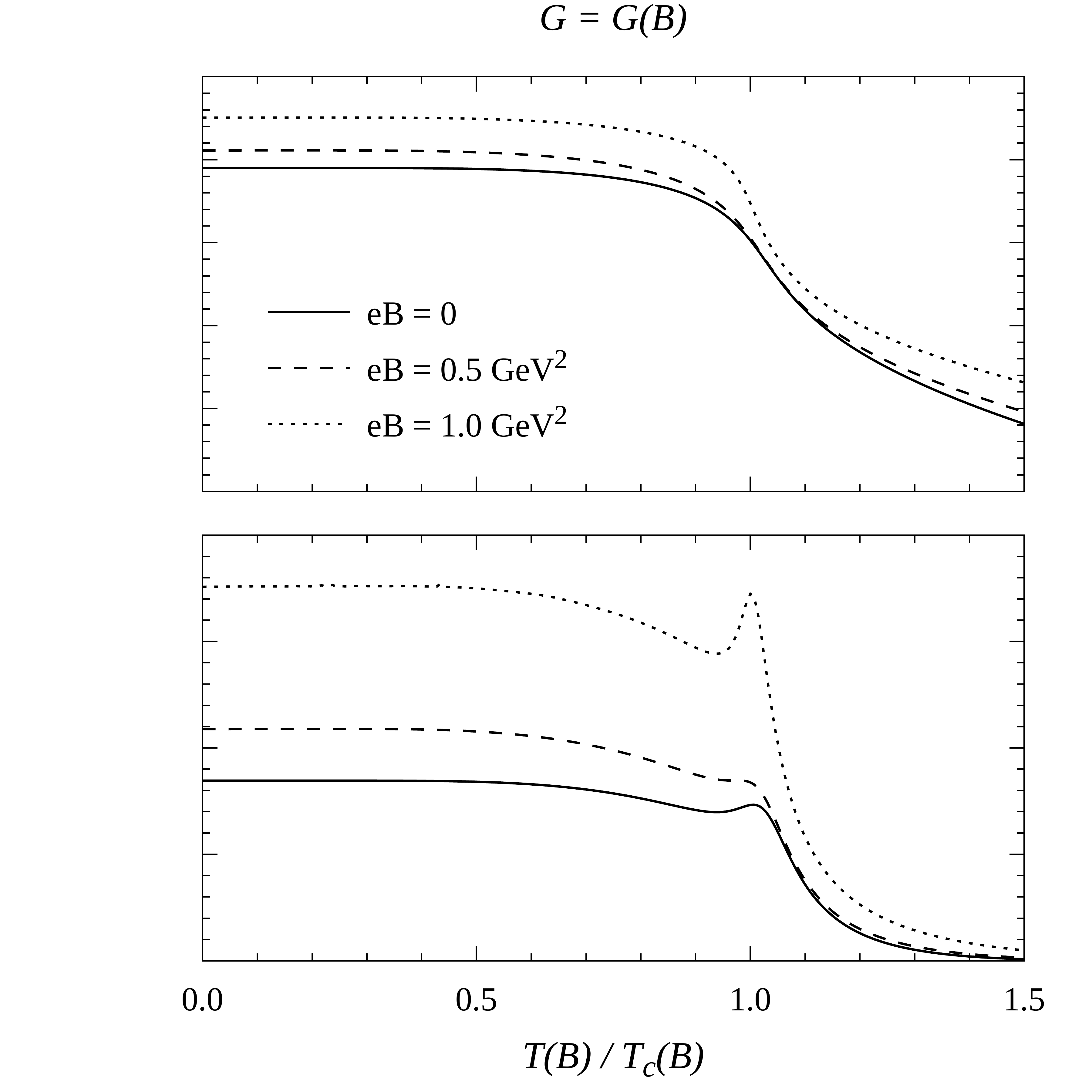}
\caption{Values of $\chi_t^{1/4}$ and $\lambda_a\,f_a^4$ as functions of
$T(B)/T_c(B)$ for some representative values of the magnetic field.}
\label{fig3}
\end{figure}

The behavior of the above quantities for nonzero temperatures is shown in
Fig.~\ref{fig3}, where we show the numerical results for $\chi_t^{1/4}$ and
$\lambda_a\,f_a^4$ as functions of $T/T_c(B)$ for three representative
values of the magnetic field. The pseudocritical chiral transition
temperatures $T_c(B)$ have been defined taking the maximum values of the
slopes $d\bar s_l/dT$, where $\bar s_l=(\bar s_u+\bar s_d)/2$, for each
value of the magnetic field. Left and right panels correspond to the results
for constant and $B$-dependent couplings, respectively. As expected, one
finds a sudden drop of both $\chi_t^{1/4}$ and $\lambda_a$ at $T=T_c$,
signalling the restoration of chiral symmetry in the light quark sector.
Notice that the curves for $\lambda_a$ tend to show a peak located at
$T=T_c$.

\begin{figure}[hbt]
\centering
\includegraphics[width=0.95\textwidth]{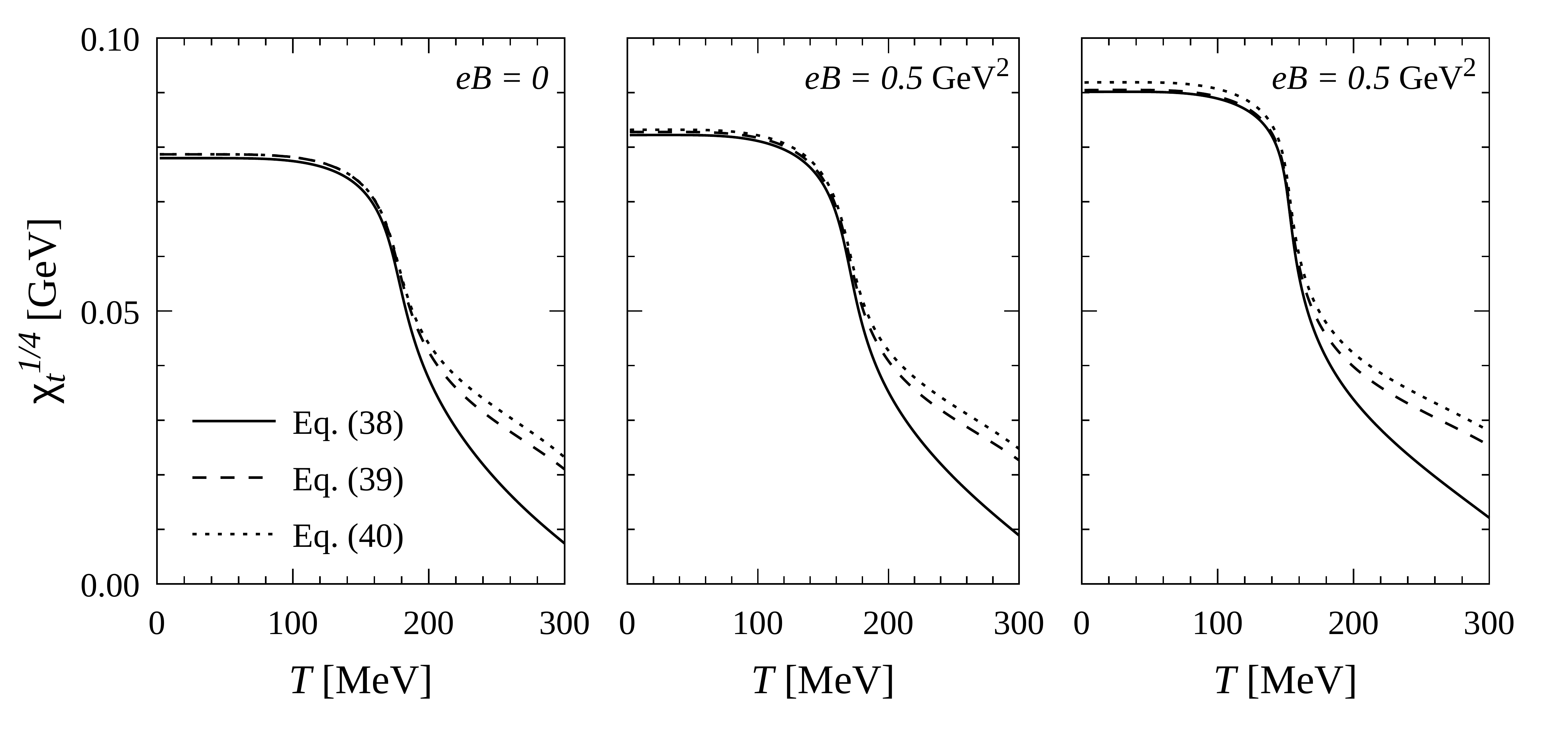}
\caption{Values of $\chi_t^{1/4}$ as calculated within our model
---Eq.~(\ref{chit_exacta})--- compared with the results arising from the
approximate expressions in Eqs.~(\ref{chit_app1}) and (\ref{chit_app2}), for
three different values of the magnetic field. The curves correspond to the
case $G=G(B)$.} \label{fig4}
\end{figure}

In Fig.~\ref{fig4} we compare the results for $\chi_t^{1/4}$ obtained within
our model, Eq.~(\ref{chit_exacta}) (solid lines) with those arising from the
approximate expressions in Eqs.~(\ref{chit_app1}) and (\ref{chit_app2})
(dashed and dotted lines, respectively). The curves correspond to the case
$G=G(B)$, for $eB=0$, 0.5~GeV$^2$ and 1~GeV$^2$. It can be seen that up to
$T\simeq T_c$ all expressions are approximately equivalent for the
considered values of the magnetic field. Beyond the chiral restoration
transition, the curves corresponding to the approximate expressions in
Eqs.~(\ref{chit_app1}) and (\ref{chit_app2}) show some deviation with
respect to the full result in Eq.~(\ref{chit_exacta}). This difference is
mainly due to the fact that the first term into the brackets, on the right
hand side of Eq.~(\ref{chit_exacta}), becomes nonnegligible in this region.

\begin{figure}[hbt]
\centering
\includegraphics[width=0.6\textwidth]{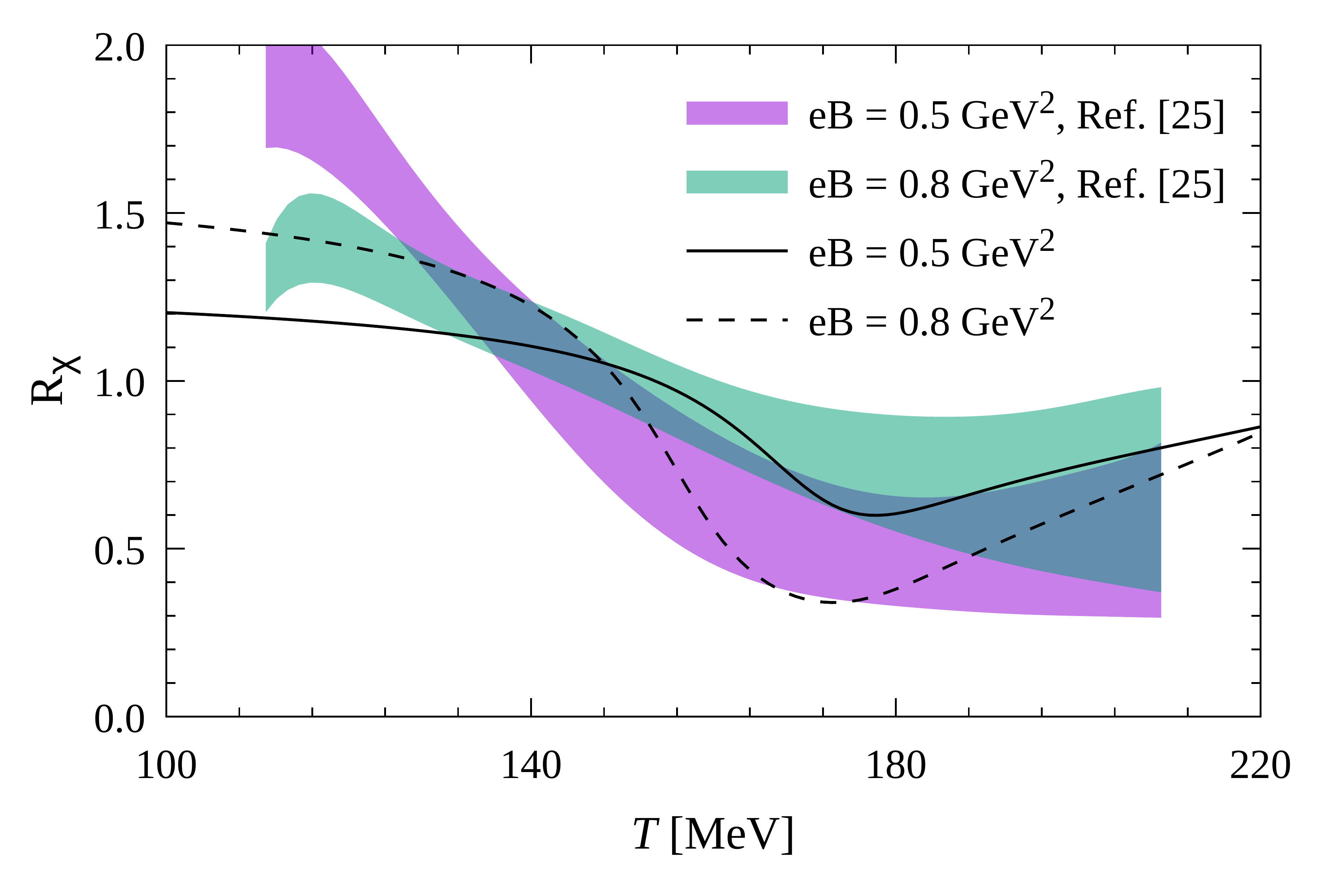}
\caption{$\chi_t(B,T)/\chi_t(0,T)$ as a function of $T$ for the case
$G=G(B)$. The shaded bands correspond to Lattice QCD results from
Ref.~\cite{Brandt:2024gso}.}
\label{fig5}
\end{figure}

Our numerical results for the temperature dependence of $\chi_t$ can also be
compared with those recently obtained from LQCD calculations, see
Ref.~\cite{Brandt:2024gso}. To perform the comparison we consider the
normalized quantity $R_\chi \equiv \chi_t(B,T)/\chi_t(0,T)$ introduced in
that work. In Fig.~\ref{fig5} we show our results (black curves) together
with those quoted in Ref.~\cite{Brandt:2024gso} (shaded bands) for $eB=0.5$
and 0.8~GeV$^2$. We include in the figure just the curves that correspond to
the case of the $B$-dependent coupling $G(B)$, which, as stated, is the one
consistent with LQCD results for IMC. Once again, it is seen that the
predictions of the NJL model show qualitative agreement with LQCD
calculations.

\begin{figure}[hbt]
\centering
\includegraphics[width=0.49\textwidth]{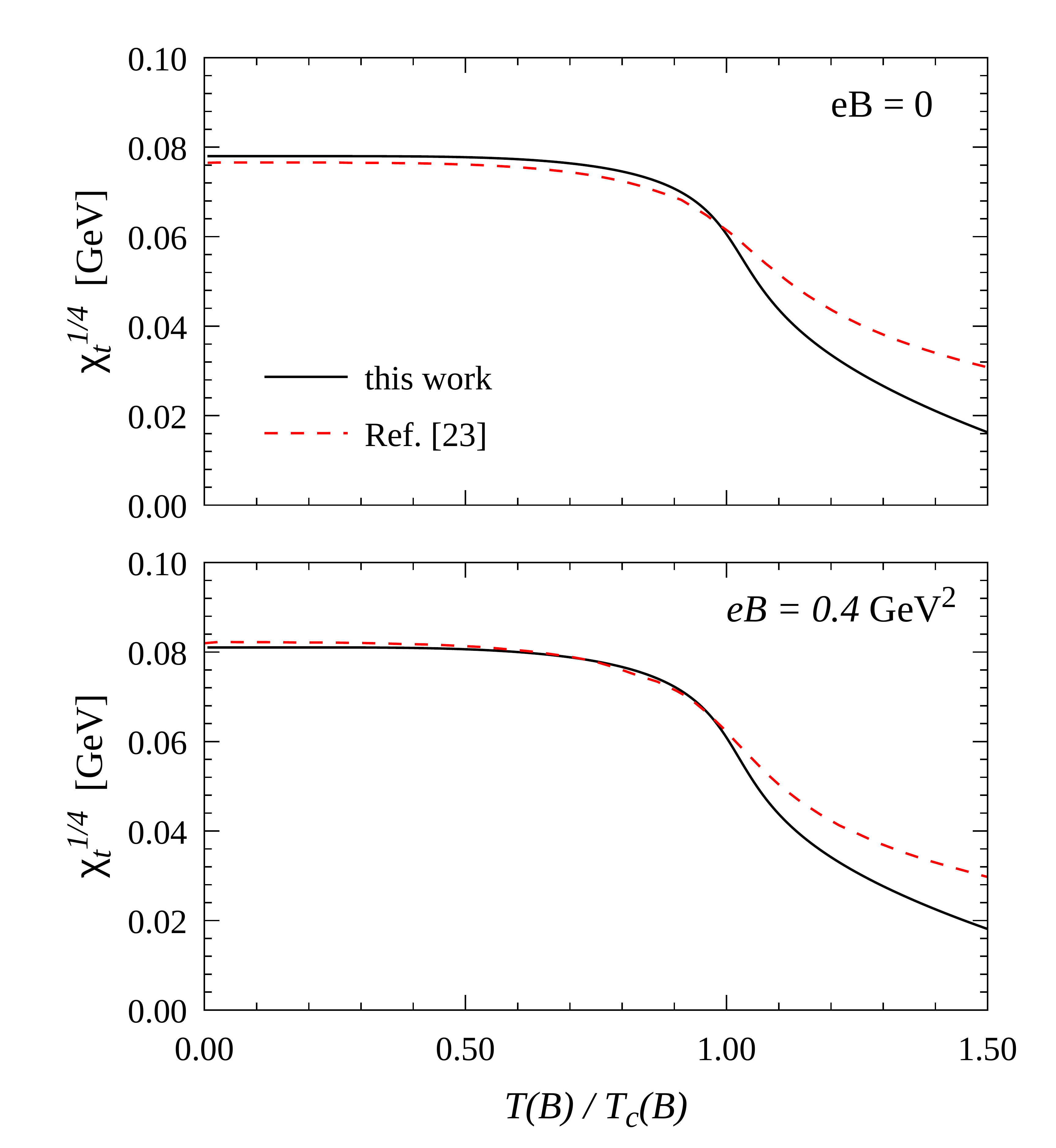}
\hspace*{-0.75cm}
\includegraphics[width=0.49\textwidth]{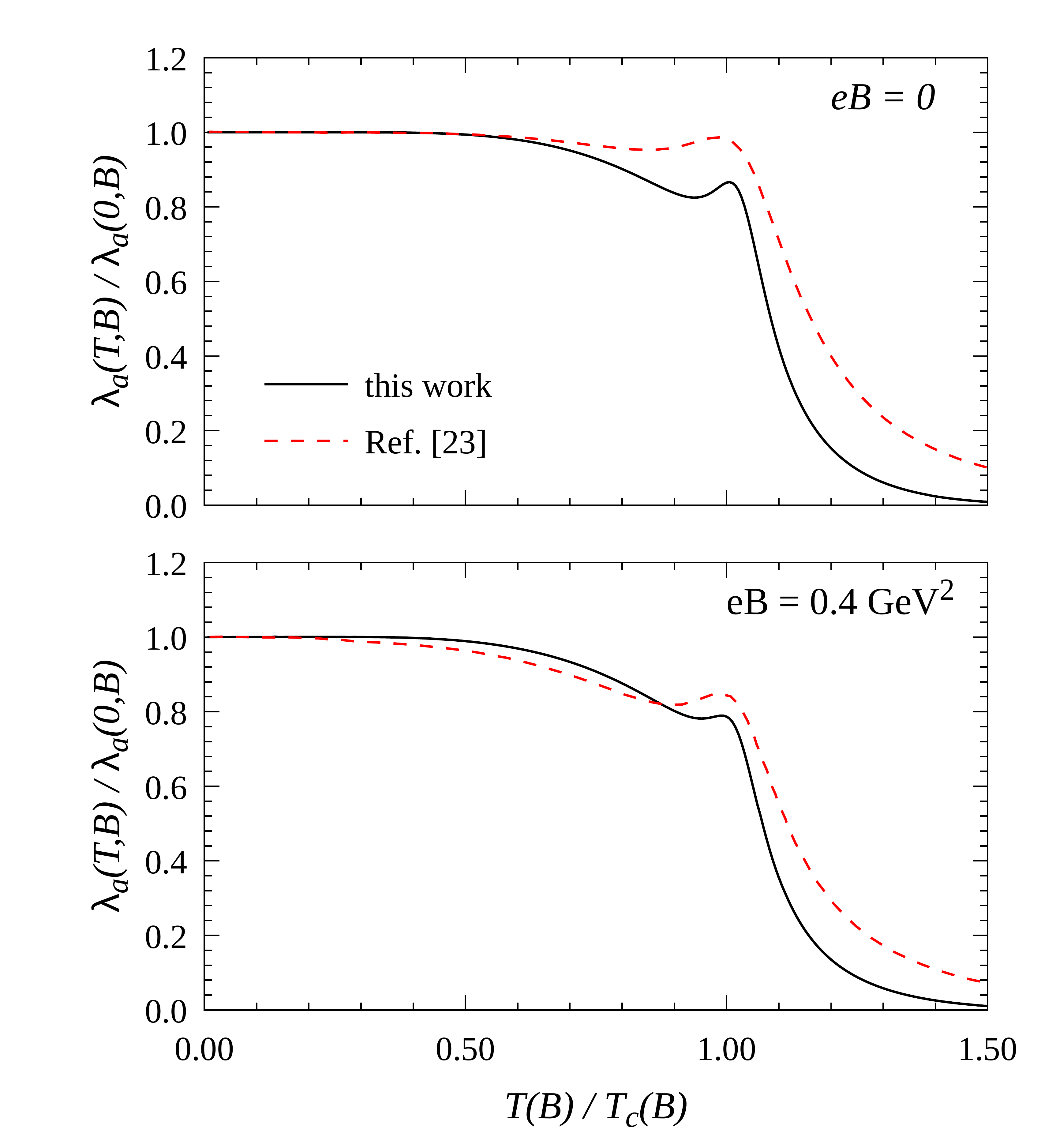}
\caption{$\chi^{1/4}(B,T)$ (left) and $\lambda_a(T,B)/\lambda_a(0,B)$ (right) as
functions of $T$ at $\mu=0$ for the case of $G=G(B)$. Our results (black solid
lines) are compared with those obtained within a two-flavor NJL model in
Ref.~\cite{Bandyopadhyay:2019pml} (red dashed lines).}
\label{fig6}
\end{figure}

In Fig.~\ref{fig6} we go back to our results for $\chi_t^{1/4}$ and
$\lambda_a$ as functions of the temperature, this time normalizing
$\lambda_a$ to the corresponding value at $T=0$ and taking the magnetic
field values $eB=0$ and $eB=0.4$~GeV$^2$, in order to compare our results
(black solid lines) with those obtained within the two-flavor NJL model
studied in Ref.~\cite{Bandyopadhyay:2019pml} (red dashed lines). As in
Fig.~\ref{fig5}, our results correspond to $G=G(B)$, given by
Eq.~(\ref{gdeb}). The values of the temperature have been normalized to the
critical temperatures $T_c(B)$, which are somewhat different for both
models. This is in part due to the fact that in
Ref.~\cite{Bandyopadhyay:2019pml} an explicit dependence on both $B$ and $T$
has been assumed for the coupling $G$. It is seen that for the two-flavor
model the peak of $\lambda_a$ at $T=T_c$ is slightly higher, while the fall
of both $\chi_t^{1/4}$ and $\lambda_a$ for $T>T_c$ is less pronounced than
in the case of the three-flavor NJL model. Nonetheless, it could be said
that the behavior of $\chi_t^{1/4}$ and $\lambda_a$ is found to be
qualitatively similar for both models.

\subsection{Finite chemical potential}

\begin{figure}[hbt]
\centering
\includegraphics[width=0.49\textwidth]{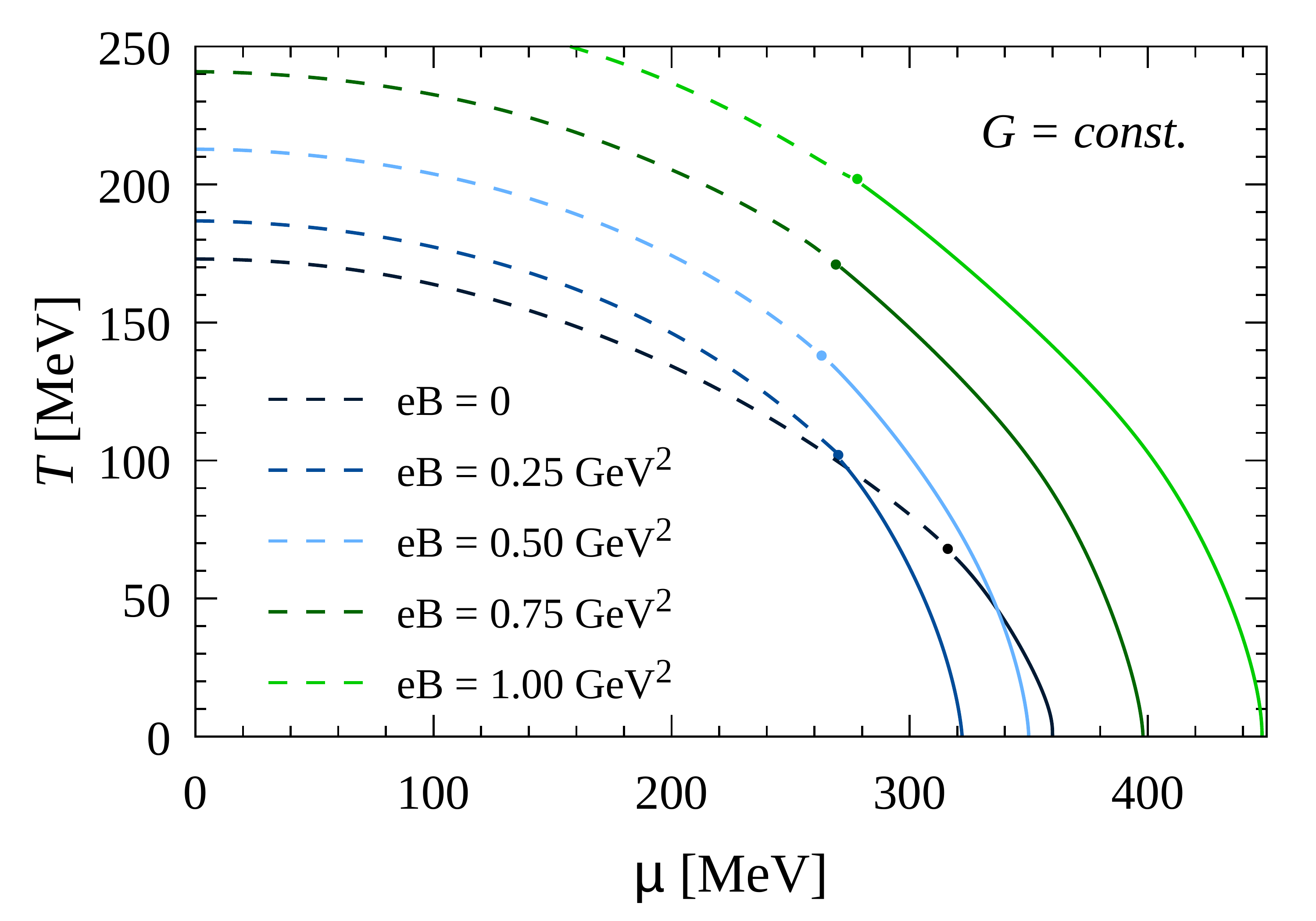}
\hspace*{-1.25cm}
\includegraphics[width=0.49\textwidth]{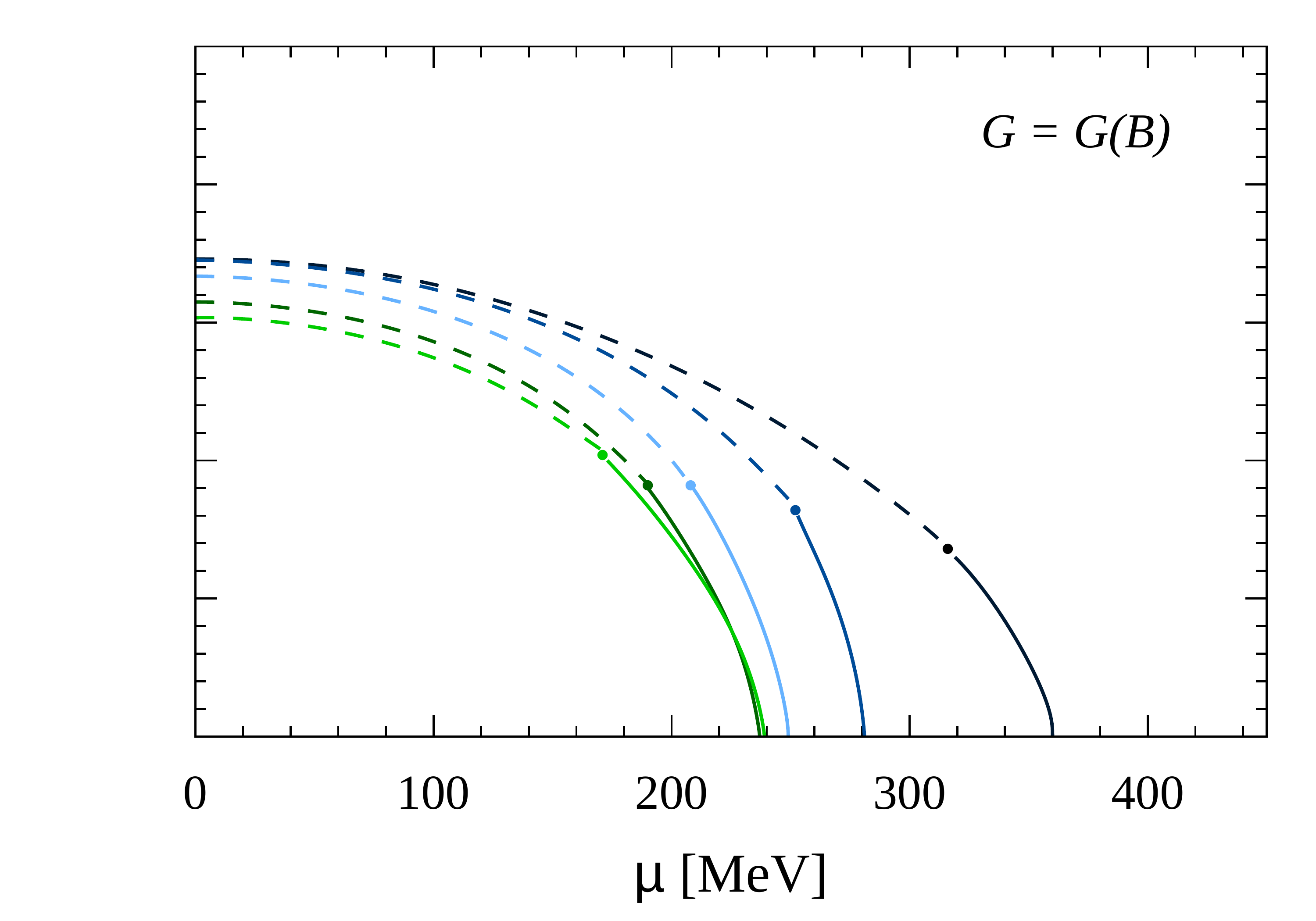}
\caption{$\mu-T$ phase diagrams for several values of the magnetic field.
Solid (dashed) lines correspond to first order (crossover) transitions,
while critical end points are indicated by the fat dots. Left and right
panels correspond to $G={\rm constant}$ and $G=G(B)$, respectively.}
\label{fig7}
\end{figure}

We turn now to discuss our numerical results for systems at nonzero quark
chemical potential $\mu$. For a better comprehension, we start by briefly
reviewing the phase diagrams in the $\mu-T$ plane, shown in Fig.~\ref{fig7}.
As expected, for low temperatures the system undergoes a first order chiral
restoration transition at given critical chemical potentials $\mu_c(B,T)$.
In the figure we show the corresponding transition lines (solid lines in the
figure) for some representative values of the magnetic field. These first
order transition lines finish at some critical end points (CEPs), whose
positions depend on the external magnetic field. For higher values of the
temperature, the transitions turn into smooth crossovers (dashed lines in
the figure), as discussed for the systems at $\mu=0$.

Left and right panels of Fig.~\ref{fig7} correspond to constant and
$B$-dependent $G$, respectively. As stated, the assumption of a $B$
dependence as the one in Eq.~(\ref{gdeb}) leads to IMC, implying a
significant change in the phase diagrams with respect to those obtained for
the $G={\rm constant}$ case. On the other hand, the behavior of the CEP with
the magnetic field is known to be strongly model-dependent. It is seen that
our results for the case $G=G(B)$ are found to be qualitatively similar to
those obtained in Ref.~\cite{Costa:2015bza}, where a three-flavor
Polyakov-NJL model with a $B$-dependent coupling is studied; however, the
behavior of the CEP is shown to be different from the one found in
Ref.~\cite{Carlomagno:2023clk}, where a nonlocal NJL-like model is
considered (notice that in this type of model IMC is naturally
obtained~\cite{Pagura:2016pwr,GomezDumm:2017iex}).

\begin{figure}[hbt]
\centering
\includegraphics[width=0.49\textwidth]{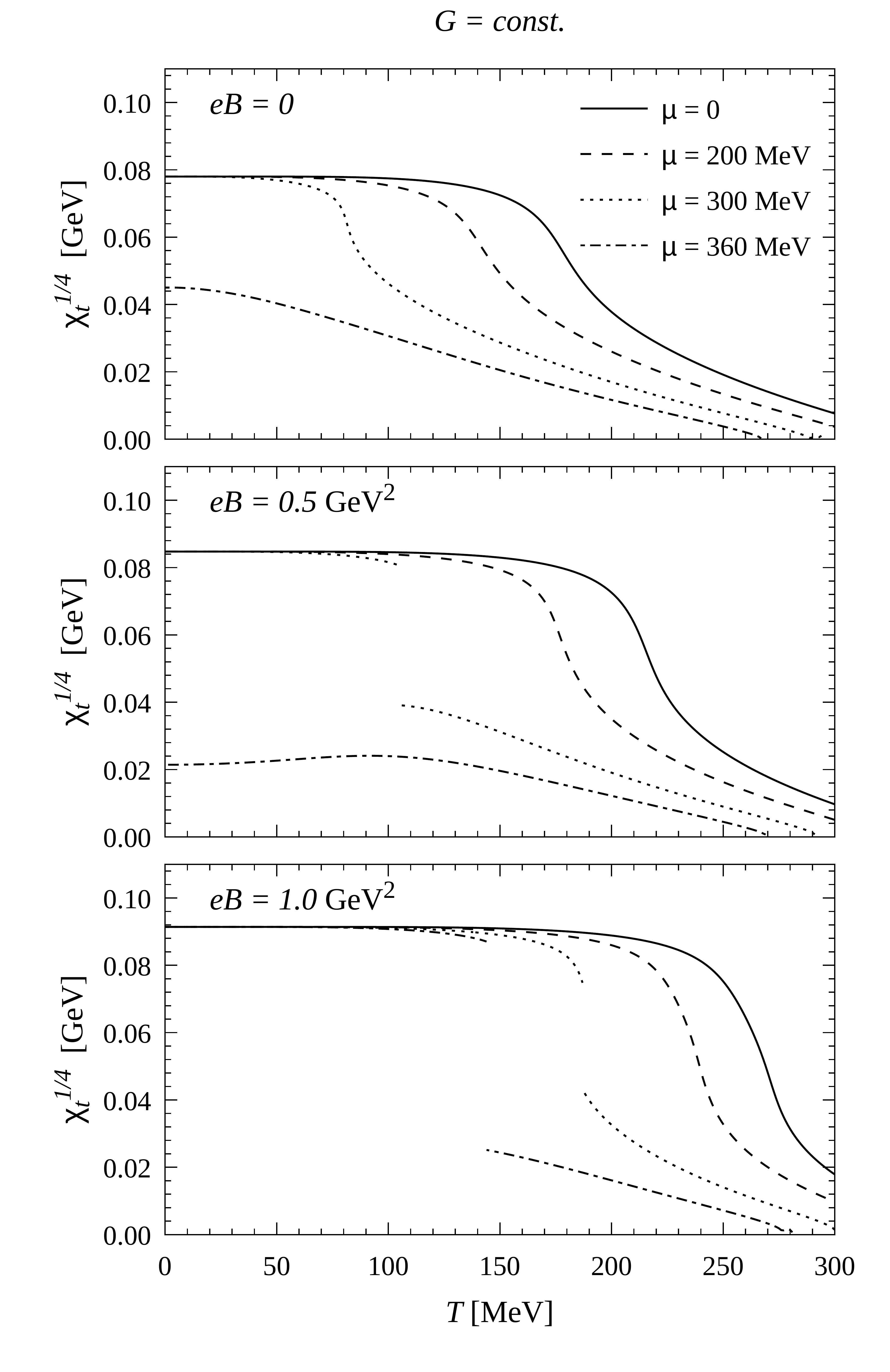}
\hspace*{-1.5cm}
\includegraphics[width=0.49\textwidth]{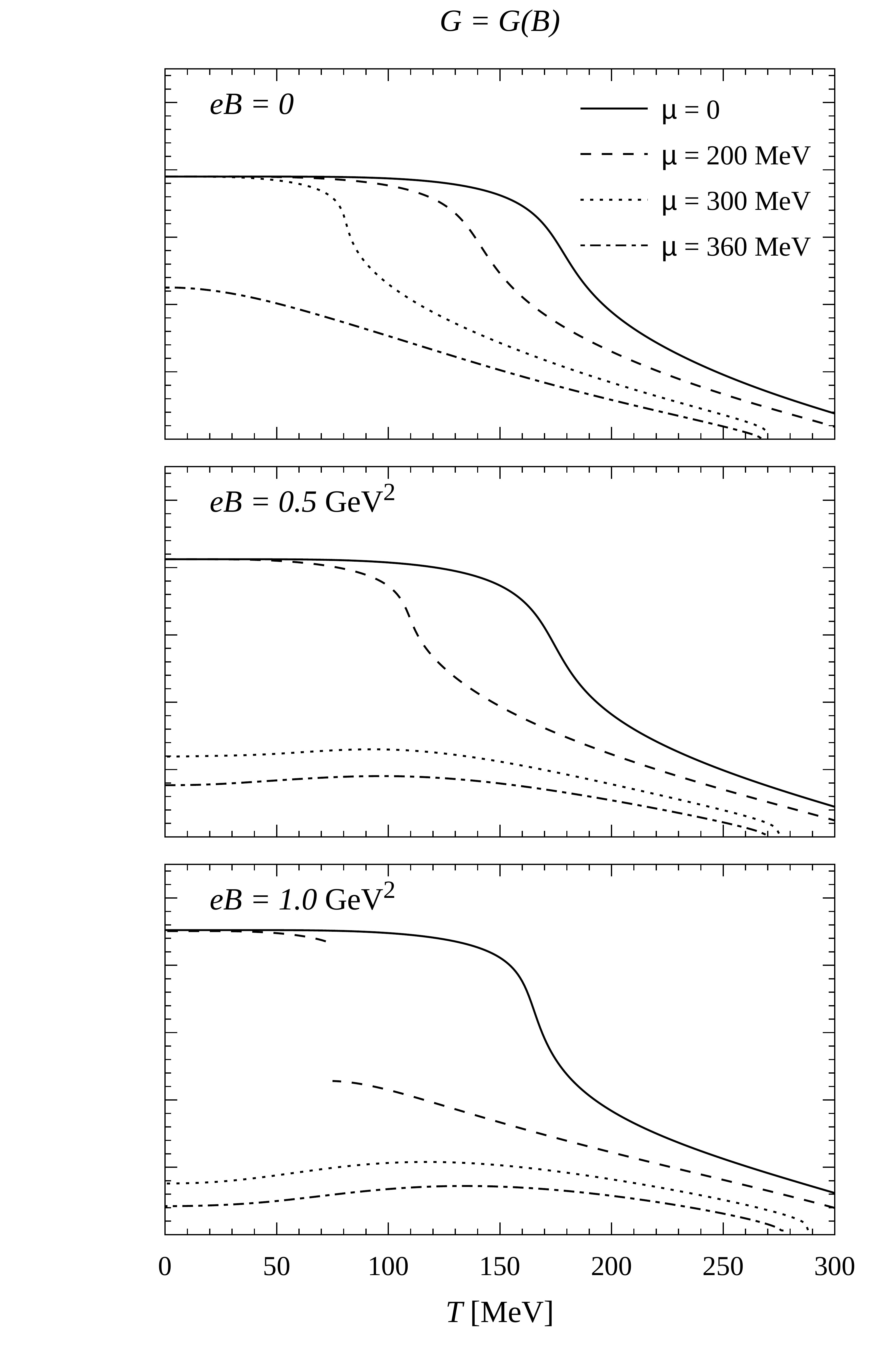}
\caption{$\chi_t^{1/4}$ as a function of $T$ for various values of the quark
chemical potential and the magnetic field. Left and right panels correspond
to $G={\rm constant}$ and $G=G(B)$, respectively.}
\label{fig8}
\end{figure}

\begin{figure}[hbt]
\centering
\includegraphics[width=0.49\textwidth]{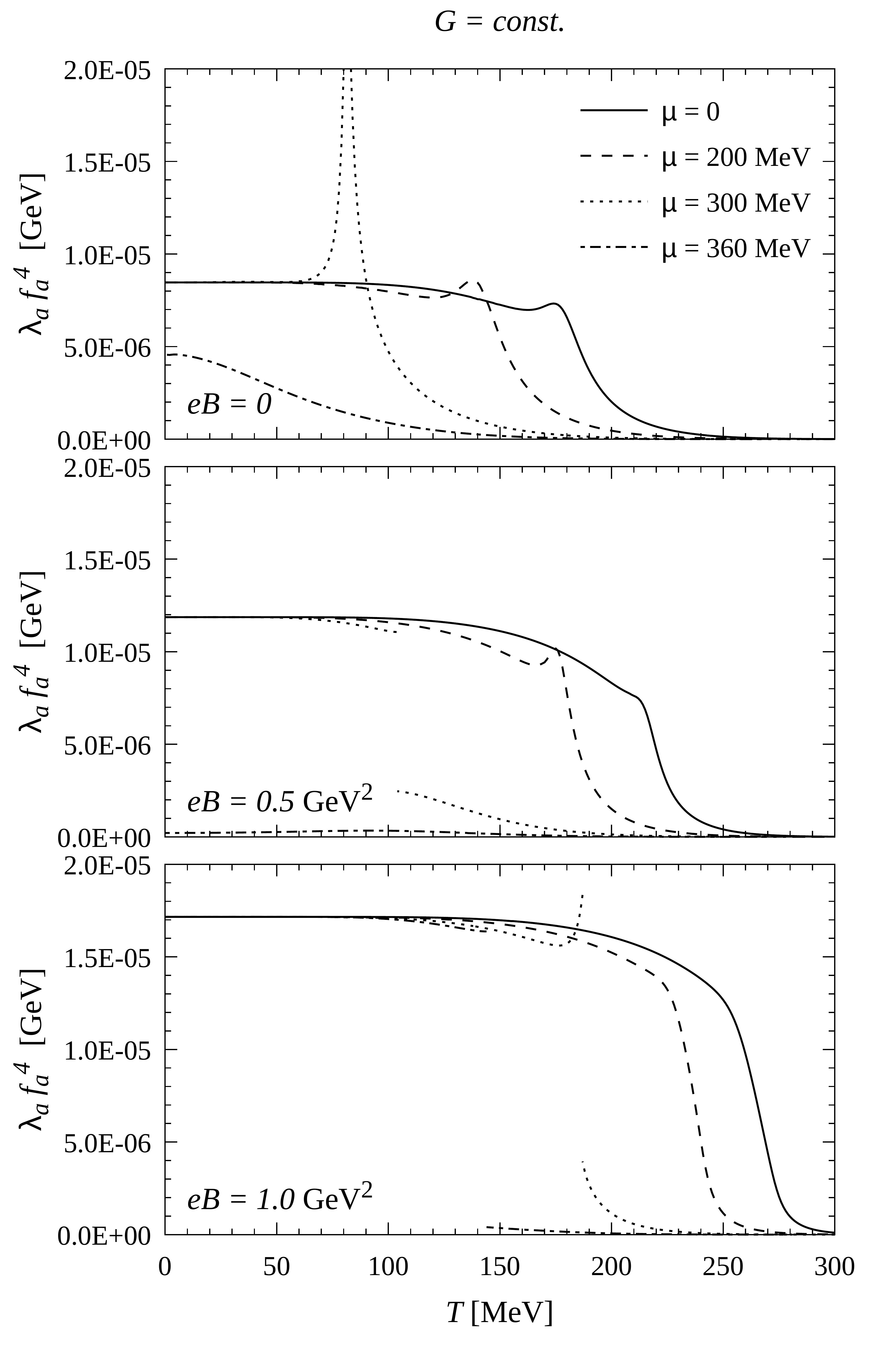}
\hspace*{-1.5cm}
\includegraphics[width=0.49\textwidth]{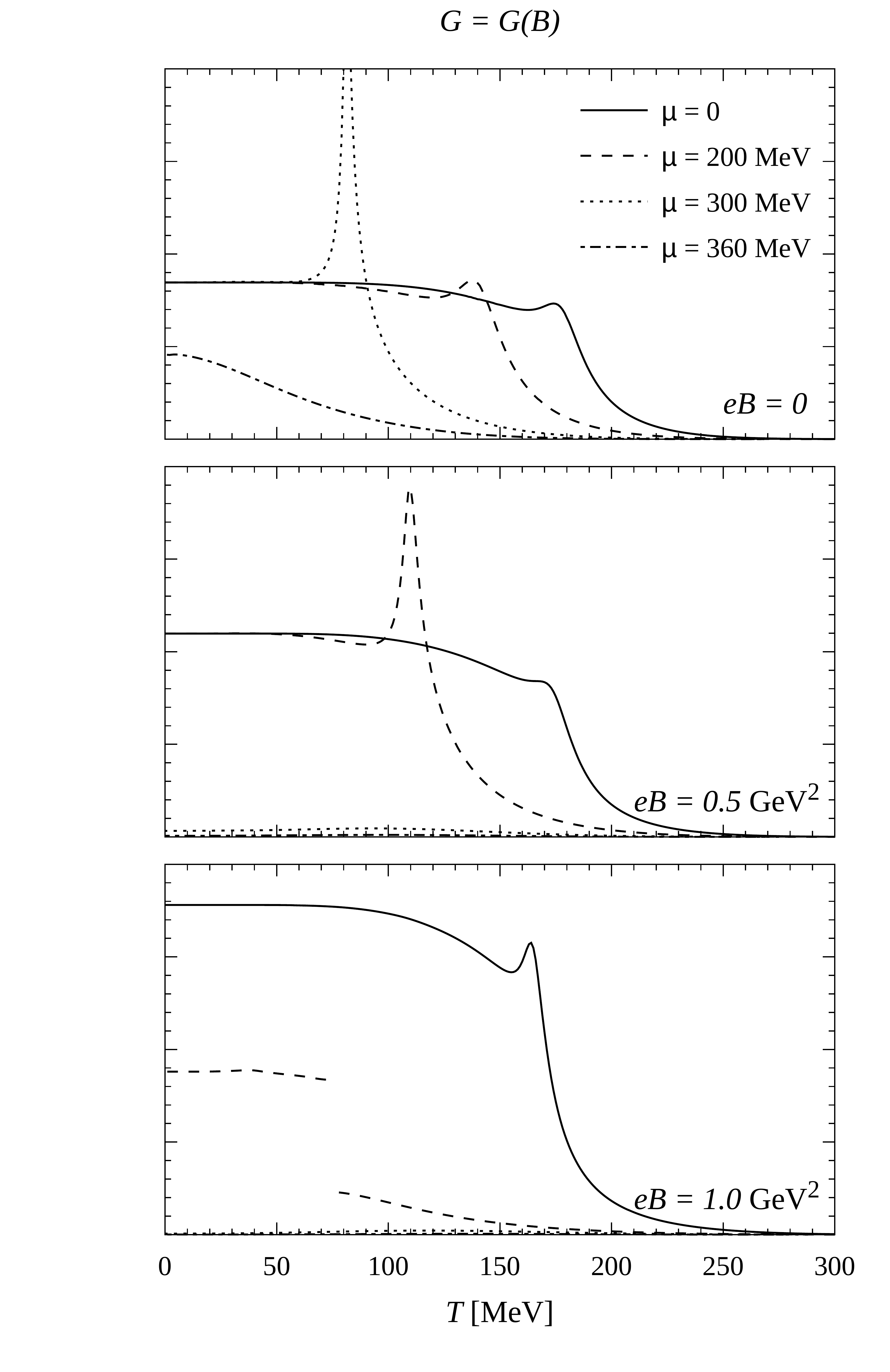}
\caption{$\lambda_a\,f_a^4$ as a function of $T$ for various values of the
quark chemical potential and the magnetic field. Left and right panels
correspond to $G={\rm constant}$ and $G=G(B)$, respectively.}
\label{fig9}
\end{figure}

In Fig.~\ref{fig8} we show the behavior of $\chi_t^{1/4}$ as a function of
the temperature, taking some representative values of the chemical potential
and the magnetic field. The curves show clearly the first order and
crossover transitions, both for the cases of $G ={\rm constant}$ (left
panels) and $G=G(B)$ (right panels). We have also verified that
for nonzero $\mu$ the expressions in Eqs.~(\ref{chit_app1}) and
(\ref{chit_app2}) still approximate with good accuracy ($\lesssim 1$\%) the
exact result in Eq.~(\ref{chit_exacta}), for temperatures that lie below the
chiral transition. Then, in Fig.~\ref{fig9} we show the behavior of the
axion self-coupling $\lambda_a$ (times the dimensionful scale $f_a^4$), for
the same values of $\mu$ and $eB$. It is seen that the peak in $\lambda_a$
---which at $\mu=0$ was found to occur at the pseudocritical temperature---
goes to infinity when the transition reaches the CEP, and turns into a first
order transition jump beyond the CEP. This shows that at the transition the
coupling $\lambda_a$ behaves like the derivative of an order parameter, and
the position of the peak could be used to define the pseudocritical
transition temperature related with the restoration of the U(1)$_A$
symmetry.

\begin{figure}[hbt]
\centering
\includegraphics[width=0.55\textwidth]{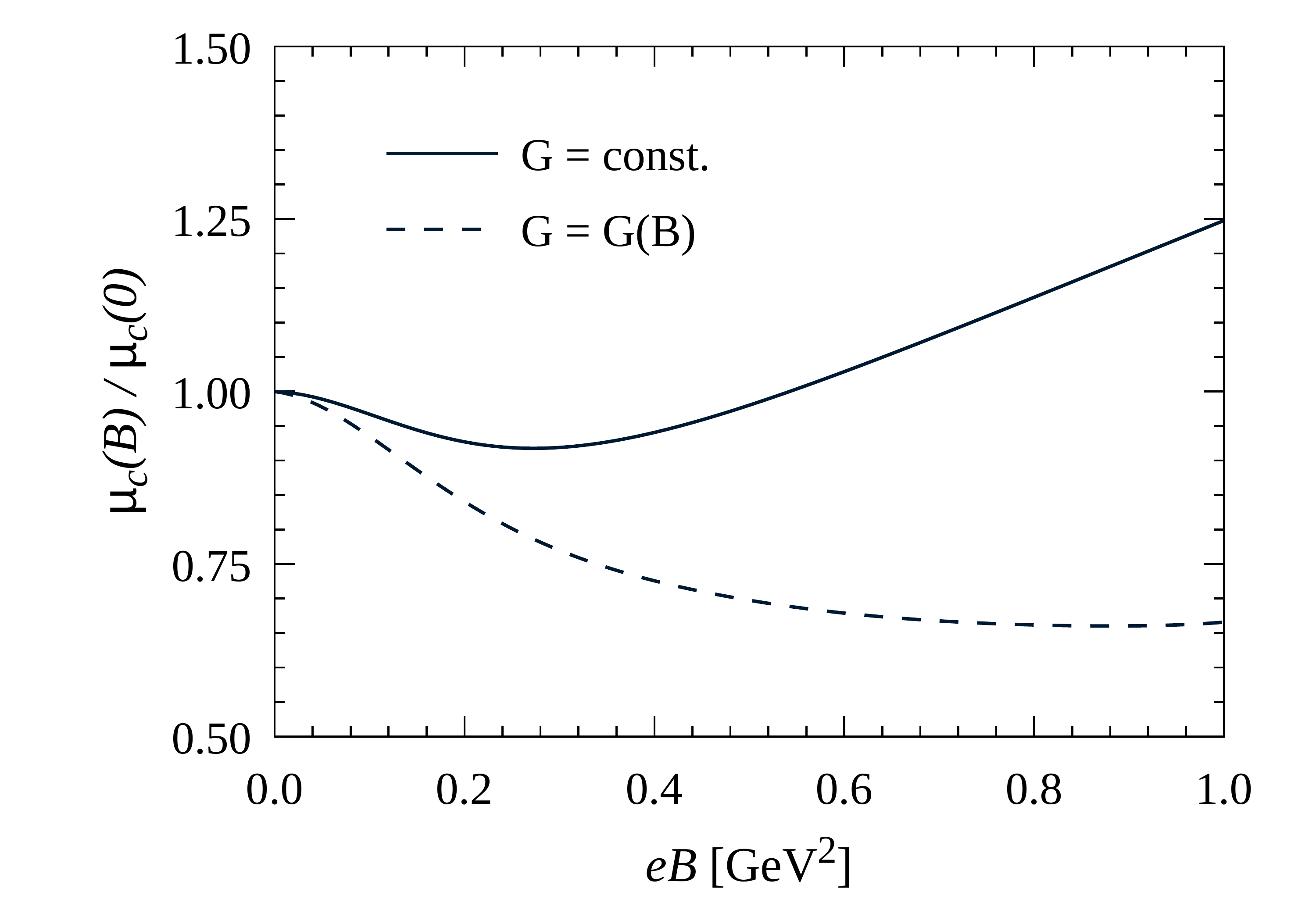}
\caption{Critical chemical potentials as functions of $eB$, for zero
temperature. Values are normalized to $\mu_c(B=0)=359$~MeV. Solid and dashed
lines correspond to $G={\rm constant}$ and $G=G(B)$, respectively.}
\label{fig10}
\end{figure}

To conclude this section, we discuss the numerical results obtained for zero
temperature and finite quark chemical potential. In Fig.~\ref{fig10} we show
the behavior of the critical chemical potential $\mu_c(B,0)$ as a function
of the magnetic field, both for constant and $B$-dependent couplings. The
values are normalized to $\mu_c(0,0)$. It is interesting to notice that for
$B\lesssim 0.3$~GeV$^2$ the models show again an IMC-like effect, i.e., a
decrease of the critical chemical potential for increasing values of the
magnetic field. This effect has been observed in various effective
approaches to low energy QCD, including local and nonlocal NJL-like
models~\cite{Preis:2010cq,Preis:2012fh,Allen:2013lda,Allen:2015paa,Ferraris:2021vun}.
For $G=G(B)$ it is seen that the IMC extends to larger values of the
magnetic field, while for constant $G$ the values of $\mu_c$ reach a minimum
and then get increased. This growth is consistent with the results in
Ref.~\cite{Allen:2013lda}, where a constant value of $G$ is assumed, while
the persistent IMC behavior is similar to the one obtained in nonlocal NJL
models~\cite{Ferraris:2021vun}, where nonlocality leads to an effective
dependence of the couplings on the magnetic field~\cite{GomezDumm:2017iex}.

\begin{figure}[H]
\centering
\includegraphics[width=0.49\textwidth]{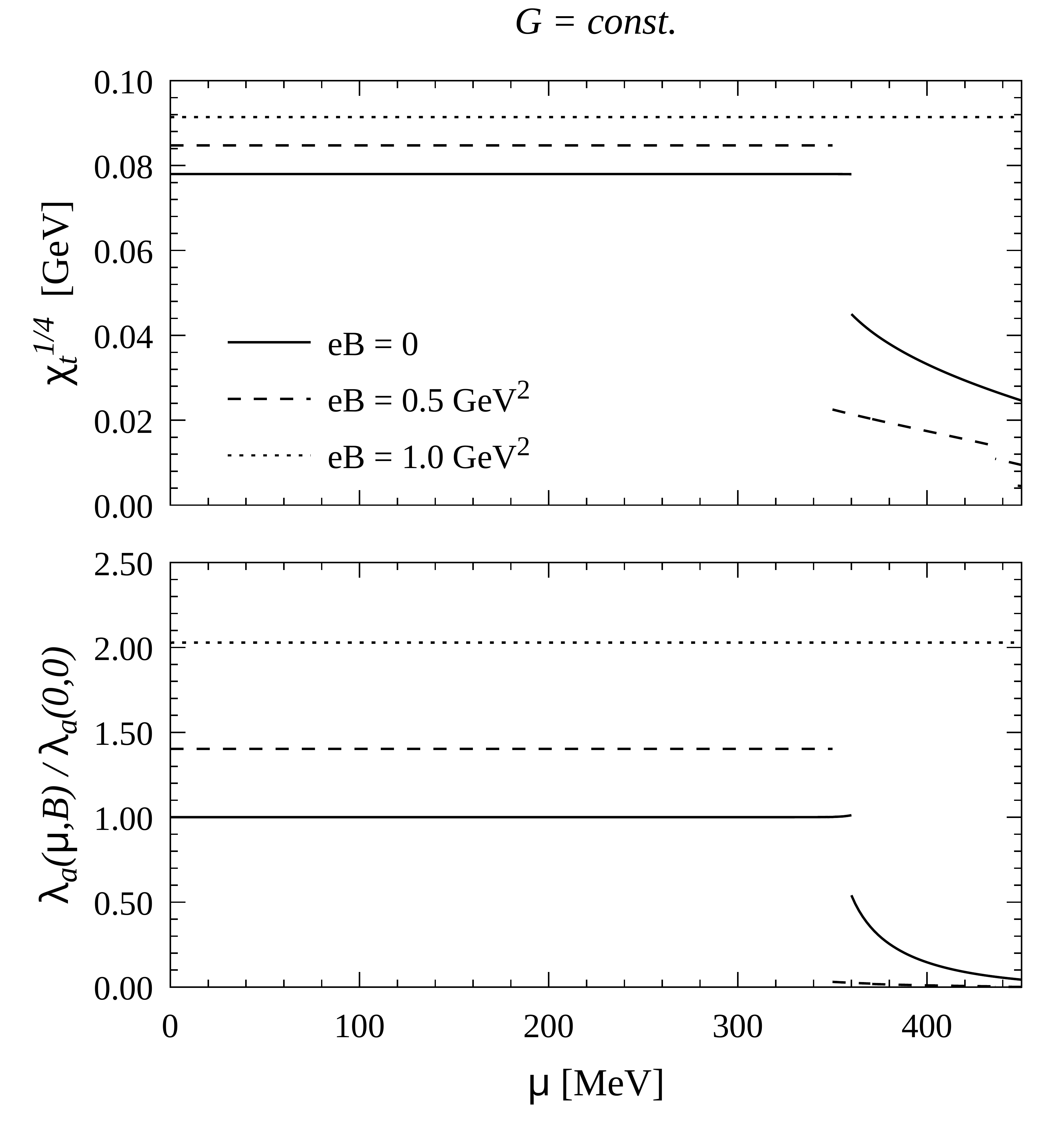}
\hspace*{-1.25cm}
\includegraphics[width=0.49\textwidth]{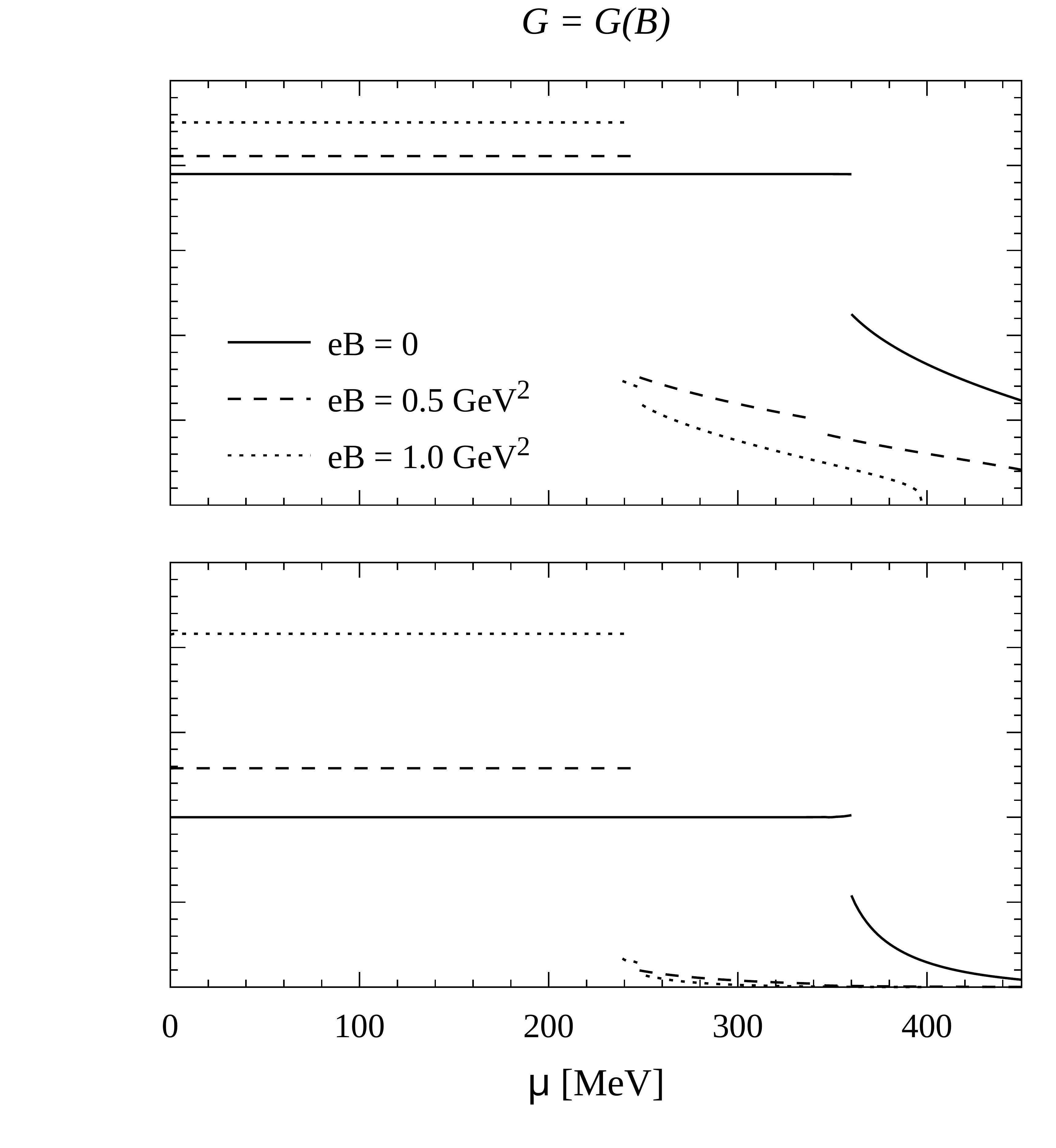}
\caption{Values of $\chi_t^{1/4}$ and $\lambda_a(\mu,B)/\lambda_a(0,0)$ as
functions of $\mu$ for some representative values of the magnetic field. Left
and right panels correspond to $G = {\rm constant}$ and $G = G(B)$, respectively.}
\label{fig11}
\end{figure}

\begin{figure}[hbt]
\centering
\includegraphics[width=0.6\textwidth]{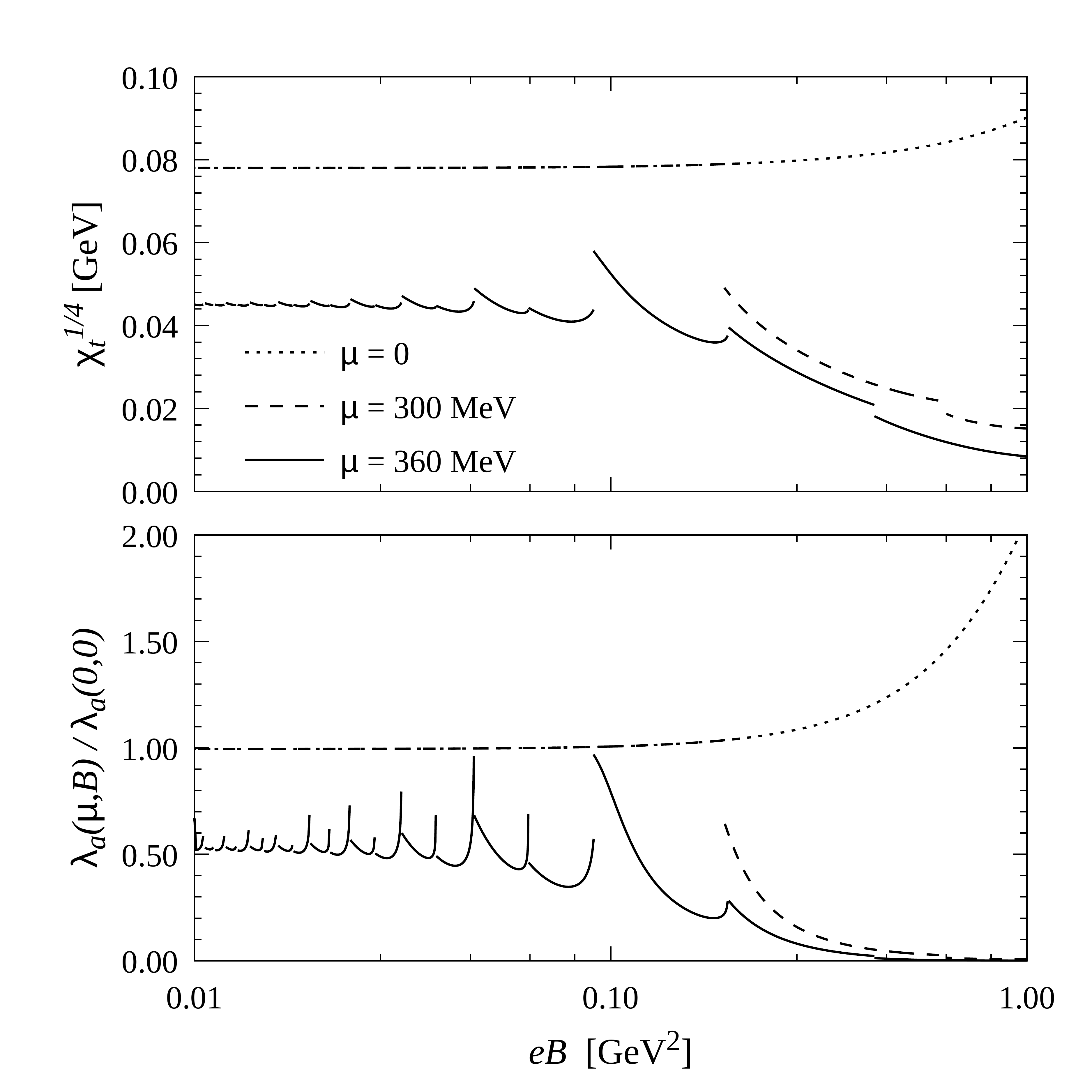}
\caption{Values of $\chi^{1/4}$ and $\lambda_a(\mu,B)/\lambda_a(0,0)$ as
functions of $eB$ for some representative values of the quark chemical
potential. The curves correspond to the case $G = G(B)$.}
\label{fig12}
\end{figure}

In Fig.~\ref{fig11} we show the behavior of $\chi_t^{1/4}$ and
$\lambda_a/\lambda_a(\mu=0,B=0)$ as functions of the chemical potential, for
three representative values of $eB$. Left and right panels correspond to
constant and $B$-dependent couplings, respectively. The first order chiral
transitions can be clearly observed. Moreover, it can be seen that beyond
these transitions there is a second discontinuity, which corresponds to the
partial chiral symmetry restoration related to the $s$ quark-antiquark
condensate. Finally, in Fig.~\ref{fig12} we show the behavior of
$\chi_t^{1/4}$ and $\lambda_a/\lambda_a(\mu=0,B=0)$ as functions of the
magnetic field, for some selected values of $\mu$. Here we use a logarithmic
scale, in order to focus on the region of low $eB$, and we just include the
results for $G=G(B)$ (the curves for $G={\rm constant}$ are similar for
values of $eB$ up to about 0.3~GeV$^2$). As one can see from
Fig.~\ref{fig10}, for low values of $\mu$ the system lies in the chiral
symmetry broken phase for all considered values of $eB$. Then, for
$\mu\gtrsim 230$~MeV a first order transition is found at some intermediate
value of $eB$, while for values of $\mu$ beyond $\mu_c(0,0) = 359$~MeV the
system lies in the partially restored chiral symmetry phase. Typically, in
this region one finds for $T=0$ a series of magnetic oscillations related to
the van Alphen-de Haas effect~\cite{Ebert:1999ht}. In fact, as shown in the
figure, one finds a sequence of first order transitions that correspond to
the values of $\mu$ that satisfy the relation $\mu^2 = 2 k B_f + \bar M_f^2$
($f=u,d$) for integer $k$.
The value of $\chi_t^{1/4}$ is found to be relatively large for $\mu$ right
above $\mu_c$ at low values of the magnetic field, and becomes significantly
reduced when $eB$ is increased up to $\sim 1$~GeV$^2$.

\section{Summary and conclusions}
\label{summary}

In the present work we have analyzed the topological susceptibility and the
axion properties in the presence of a strong magnetic field, considering a
three flavor NJL model that includes strong CP violation through a 't
Hooft-like flavor mixing term. The behavior of the relevant quantities for
systems at finite temperature and quark chemical potential have been
studied.

As is well known, when the scalar/pseudoscalar coupling $G$ is kept constant
(i.e., when it does not depend on the magnetic field) local NJL models are not
able to reproduce the inverse magnetic catalysis effect at finite
temperature. Therefore, we have considered both the case of a constant $G$
and the one in which one assumes a $B$ dependent coupling $G=G(B)$, chosen
in such a way that the model is able to adequately reproduce the $B$ dependence of the critical chiral transition temperatures obtained
in lattice QCD.

We have shown that within these three flavor NJL models the topological
susceptibility has a rather simple expression (see Eq.~(\ref{chit_exacta}))
in terms of the quark condensates, the current quark masses and the strength
of the flavor mixing term. In addition, we have shown that close to the
chiral limit this expression reduces to the one obtained in other approaches
to non-perturbative QCD such as chiral perturbation
theory~\cite{Mao:2009sy,Adhikari:2022vqs} and the linear sigma
model~\cite{Kawaguchi:2020qvg}.

At $T=\mu=0$, using standard values of model parameters we obtain, for
vanishing external magnetic field, $\chi_t = 78$~MeV and $\lambda_a =
0.85\times 10^{-5}$~GeV$^4/f_a^4$, in reasonable agreement with values
obtained from LQCD and/or ChPT~\cite{Lu:2020rhp}. For nonzero magnetic field, in agreement
with previous analyses we find that the topological susceptibility gets
increased with $B$. Clearly, this can be understood by noticing that,
according to the previously mentioned theoretical expressions, $\chi_t$ is
approximately proportional to the quark condensates, which exhibit the well
known magnetic catalysis effect. Moreover, we find that the axion
self-coupling $\lambda_a$ also increases with the magnetic field.

For the case of both nonzero temperature and magnetic field, we find that,
as expected, $\chi_t$ and $\lambda_a$ remain approximately constant as
functions of $T$, up to the critical temperatures $T_c(B)$. Beyond these
values we find for both quantities a sudden drop, signalling the restoration
of the U(1)$_A$ symmetry in the light quark sector. We also note that the
curves for $\lambda_a$ tend to show a peak located at $T = T_c$. When
comparing our results with those of the SU(2) NJL model analyzed in
Ref.~\cite{Bandyopadhyay:2019pml}, it is seen that for the two-flavor model
the peak of $\lambda_a$ at $T = T_c$ is slightly higher, while the fall of
both $\chi_t^{1/4}$ and $\lambda_a$ observed for $T > T_c$ is less
pronounced than in the case of the three-flavor model. In any case, it could
be said that the behavior of $\chi_t^{1/4}$ and $\lambda_a$ is found to be
qualitatively similar for both models. Regarding the comparison with finite
temperature LQCD, our predictions for $R_\chi = \chi_t(B, T)/\chi_t(0, T)$
in the case of the $B$-dependent coupling $G(B)$ show qualitative agreement
with those obtained from LQCD calculations.

As stated, we have completed our analysis by considering
systems at nonzero quark chemical potential. Curves showing the behavior of
$\chi_t$ and $\lambda_a$ as functions of the temperature are given for
various values of the chemical potential, showing both crossover and first
order transitions. It is found that the approximate expressions for $\chi_t$
in Eqs.~(\ref{chit_app1}) and (\ref{chit_app2}) work very well (within
$\simeq 2$\%) in the chirally broken phase (i.e., at low temperatures and/or
chemical potentials), whereas they show some deviation in the restored
phase. In the case of crossover transitions, it is seen that $\lambda_a$
behaves like the derivative of an order parameter; thus, the position of the
corresponding peak can be used to define the pseudocritical transition
temperature associated with the restoration of $U(1)_A$ symmetry. At zero
temperature and finite $\mu$, it is seen that for $G=G(B)$ the critical
chemical potentials exhibit inverse magnetic catalysis, showing a similar
behavior as the one observed in nonlocal NJL-like models. In the chirally
restored phase, for $T=0$ and low magnetic fields we observe a pattern of
magnetic oscillations in the values of both $\lambda_a$ and $\chi_t$,
related to the well-known van Alphen-de Haas effect.

In general, it is seen that the case in which $G = G(B)$, which (by
construction) exhibits inverse magnetic catalysis at finite temperature,
provides the best agreement with results from LQCD at zero chemical
potential and yields a phase diagram that is consistent with the expected
phenomenology at finite chemical potential. We find that the behavior of
$\lambda_a(T,\mu,B)$ and $\chi_t(T,\mu,B)$ is predominantly determined by
the light quark condensates, i.e., by the chiral SU(2) symmetry restoration,
and shows reasonable agreement with other effective models, including
approximate results from chiral perturbation theory.

\section*{Acknowledgements}

This work has been supported in part by Consejo Nacional de Investigaciones
Cient\'ificas y T\'ecnicas and Agencia Nacional de Promoci\'on Cient\'ifica
y Tecnol\'ogica (Argentina), under Grants No.~PIP2022-GI-11220210100150CO,
No.~PICT20-01847, No.~PICT22-03-00799, and by the National University of La Plata (Argentina), Project No.~X960.


\begin{thebibliography}{0}

\bibitem{Belavin:1975fg}
A.~A.~Belavin, A.~M.~Polyakov, A.~S.~Schwartz and Y.~S.~Tyupkin,
Phys. Lett. B \textbf{59}, 85-87 (1975).

\bibitem{Callan:1976je}
C.~G.~Callan, Jr., R.~F.~Dashen and D.~J.~Gross,
Phys. Lett. B \textbf{63}, 334-340 (1976).

\bibitem{Peccei:1977hh}
R.~D.~Peccei and H.~R.~Quinn,
Phys. Rev. Lett. \textbf{38}, 1440-1443 (1977).

\bibitem{Peccei:2006as}
R.~D.~Peccei,
Lect. Notes Phys. \textbf{741}, 3-17 (2008)
[arXiv:hep-ph/0607268 [hep-ph]].

\bibitem{DiLuzio:2020wdo}
L.~Di Luzio, M.~Giannotti, E.~Nardi and L.~Visinelli,
Phys. Rept. \textbf{870}, 1-117 (2020)
[arXiv:2003.01100 [hep-ph]].

\bibitem{Preskill:1982cy}
J.~Preskill, M.~B.~Wise and F.~Wilczek,
Phys. Lett. B \textbf{120}, 127-132 (1983).

\bibitem{Abbott:1982af}
L.~F.~Abbott and P.~Sikivie,
Phys. Lett. B \textbf{120}, 133-136 (1983).

\bibitem{Dine:1982ah}
M.~Dine and W.~Fischler,
Phys. Lett. B \textbf{120}, 137-141 (1983).

\bibitem{Raffelt:2006cw}
G.~G.~Raffelt,
Lect. Notes Phys. \textbf{741}, 51-71 (2008)
[arXiv:hep-ph/0611350 [hep-ph]].

\bibitem{Sedrakian:2015krq}
A.~Sedrakian,
Phys. Rev. D \textbf{93}, no.6, 065044 (2016)
[arXiv:1512.07828 [astro-ph.HE]].


\bibitem{Fukushima:2001hr}
K.~Fukushima, K.~Ohnishi and K.~Ohta,
Phys. Rev. C \textbf{63}, 045203 (2001)
[arXiv:nucl-th/0101062 [nucl-th]].

\bibitem{Costa:2008dp}
P.~Costa, M.~C.~Ruivo, C.~A.~de Sousa, H.~Hansen and
W.~M.~Alberico,
Phys. Rev. D \textbf{79}, 116003 (2009)
[arXiv:0807.2134 [hep-ph]].

\bibitem{Contrera:2009hk}
G.~A.~Contrera, D.~G.~Dumm and N.~N.~Scoccola,
Phys. Rev. D \textbf{81}, 054005 (2010)
[arXiv:0911.3848 [hep-ph]].


\bibitem{Borsanyi:2016ksw}
S.~Borsanyi, Z.~Fodor, J.~Guenther, K.~H.~Kampert, S.~D.~Katz,
T.~Kawanai, T.~G.~Kovacs, S.~W.~Mages, A.~Pasztor and F.~Pittler,
\textit{et al.}
Nature \textbf{539}, no.7627, 69-71 (2016)
[arXiv:1606.07494 [hep-lat]].

\bibitem{Petreczky:2016vrs}
P.~Petreczky, H.~P.~Schadler and S.~Sharma,
Phys. Lett. B \textbf{762}, 498-505 (2016)
[arXiv:1606.03145 [hep-lat]].

\bibitem{Taniguchi:2016tjc}
Y.~Taniguchi, K.~Kanaya, H.~Suzuki and T.~Umeda,
Phys. Rev. D \textbf{95}, no.5, 054502 (2017)
[arXiv:1611.02411 [hep-lat]].

\bibitem{Fukushima:2008xe}
K.~Fukushima, D.~E.~Kharzeev and H.~J.~Warringa,
Phys. Rev. D \textbf{78}, 074033 (2008)
[arXiv:0808.3382 [hep-ph]].

\bibitem{Kharzeev:2024zzm}
D.~E.~Kharzeev, J.~Liao and P.~Tribedy,
[arXiv:2405.05427 [nucl-th]].

\bibitem{Asakawa:2010bu}
M.~Asakawa, A.~Majumder and B.~Muller,
Phys. Rev. C \textbf{81}, 064912 (2010)
[arXiv:1003.2436 [hep-ph]].


\bibitem{Adhikari:2021lbl}
P.~Adhikari,
Phys. Lett. B \textbf{825}, 136826 (2022)
[arXiv:2103.05048 [hep-ph]].

\bibitem{Adhikari:2021jff}
P.~Adhikari,
Nucl. Phys. B \textbf{974}, 115627 (2022)
[arXiv:2111.06196 [hep-ph]].

\bibitem{Adhikari:2022vqs}
P.~Adhikari,
Nucl. Phys. B \textbf{982}, 115853 (2022)
[arXiv:2203.00200 [hep-ph]].


\bibitem{Bandyopadhyay:2019pml}
A.~Bandyopadhyay, R.~L.~S.~Farias, B.~S.~Lopes and R.~O.~Ramos,
Phys. Rev. D \textbf{100}, no.7, 076021 (2019)
arXiv:1906.09250 [hep-ph]].

\bibitem{Ali:2020jsy}
M.~S.~Ali, C.~A.~Islam and R.~Sharma,
Phys. Rev. D \textbf{104}, no.11, 114026 (2021)
[arXiv:2009.13563 [hep-ph]].

\bibitem{Brandt:2024gso}
B.~B.~Brandt, G.~Endr\H{o}di, J.~J.~H.~Hern\'andez and G.~Mark\'o,
[arXiv:2409.00796 [hep-lat]].

\bibitem{Chatterjee:2014csa}
B.~Chatterjee, H.~Mishra and A.~Mishra,
Phys. Rev. D \textbf{91}, no.3, 034031 (2015)
[arXiv:1409.3454 [hep-ph]].

\bibitem{Boomsma:2009eh}
J.~K.~Boomsma and D.~Boer,
Phys. Rev. D \textbf{80}, 034019 (2009)
[arXiv:0905.4660 [hep-ph]].

\bibitem{Miransky:2015ava}
V.~A.~Miransky and I.~A.~Shovkovy,
Phys. Rept. \textbf{576}, 1-209 (2015)
[arXiv:1503.00732 [hep-ph]].

\bibitem{Andersen:2014xxa}
J.~O.~Andersen, W.~R.~Naylor and A.~Tranberg,
Rev. Mod. Phys. \textbf{88}, 025001 (2016)
[arXiv:1411.7176 [hep-ph]].

\bibitem{Mao:2009sy}
Y.~Y.~Mao \textit{et al.} [TWQCD],
Phys. Rev. D \textbf{80}, 034502 (2009)
[arXiv:0903.2146 [hep-lat]].

\bibitem{Kawaguchi:2020qvg}
M.~Kawaguchi, S.~Matsuzaki and A.~Tomiya,
Phys. Rev. D \textbf{103}, no.5, 054034 (2021)
[arXiv:2005.07003 [hep-ph]].

\bibitem{Leutwyler:1992yt}
H.~Leutwyler and A.~V.~Smilga,
Phys. Rev. D \textbf{46}, 5607-5632 (1992).

\bibitem{Rehberg:1995kh}
P.~Rehberg, S.~P.~Klevansky and J.~Hufner,
Phys. Rev. C \textbf{53} (1996), 410-429
[arXiv:hep-ph/9506436 [hep-ph]].

\bibitem{Coppola:2024uvz}
M.~Coppola, W.~R.~Tavares, S.~S.~Avancini, J.~C.~Sodr\'e and N.~N.~Scoccola,
[arXiv:2410.05568 [hep-ph]].

\bibitem{Ferreira:2014kpa}
M.~Ferreira, P.~Costa, O.~Louren\c{c}o, T.~Frederico and C.~Provid\^encia,
Phys. Rev. D \textbf{89} (2014) no.11, 116011
[arXiv:1404.5577 [hep-ph]].

\bibitem{Farias:2014eca}
R.~L.~S.~Farias, K.~P.~Gomes, G.~I.~Krein and M.~B.~Pinto,
Phys. Rev. C \textbf{90} (2014) no.2, 025203
[arXiv:1404.3931 [hep-ph]].

\bibitem{Farias:2016gmy}
R.~L.~S.~Farias, V.~S.~Timoteo, S.~S.~Avancini, M.~B.~Pinto and G.~Krein,
Eur. Phys. J. A \textbf{53} (2017) no.5, 101
[arXiv:1603.03847 [hep-ph]].

\bibitem{Bali:2011qj}
G.~S.~Bali, F.~Bruckmann, G.~Endrodi, Z.~Fodor, S.~D.~Katz, S.~Krieg, A.~Schafer and K.~K.~Szabo,
JHEP \textbf{02} (2012), 044
[arXiv:1111.4956 [hep-lat]].

\bibitem{HotQCD:2018pds}
A.~Bazavov \textit{et al.} [HotQCD],
Phys. Lett. B \textbf{795}, 15-21 (2019)
[arXiv:1812.08235 [hep-lat]].

\bibitem{Gorghetto:2018ocs}
M.~Gorghetto and G.~Villadoro,
JHEP \textbf{03}, 033 (2019)
[arXiv:1812.01008 [hep-ph]].

\bibitem{GrillidiCortona:2015jxo}
G.~Grilli di Cortona, E.~Hardy, J.~Pardo Vega and G.~Villadoro,
JHEP \textbf{01}, 034 (2016)
[arXiv:1511.02867 [hep-ph]].

\bibitem{Costa:2015bza}
P.~Costa, M.~Ferreira, D.~P.~Menezes, J.~Moreira and C.~Provid\^encia,
Phys. Rev. D \textbf{92}, no.3, 036012 (2015)
[arXiv:1508.07870 [hep-ph]].

\bibitem{Carlomagno:2023clk}
J.~P.~Carlomagno, S.~A.~Ferraris, D.~Gomez Dumm and A.~G.~Grunfeld,
Phys. Rev. D \textbf{108}, no.5, 056029 (2023)
[arXiv:2305.15540 [hep-ph]].

\bibitem{Pagura:2016pwr}
V.~P.~Pagura, D.~Gomez Dumm, S.~Noguera and N.~N.~Scoccola,
Phys. Rev. D \textbf{95}, no.3, 034013 (2017)
[arXiv:1609.02025 [hep-ph]].

\bibitem{GomezDumm:2017iex}
D.~G\'omez Dumm, M.~F.~Izzo Villafa\~ne, S.~Noguera, V.~P.~Pagura and N.~N.~Scoccola,
Phys. Rev. D \textbf{96}, no.11, 114012 (2017)
[arXiv:1709.04742 [hep-ph]].

\bibitem{Preis:2010cq}
F.~Preis, A.~Rebhan and A.~Schmitt,
JHEP \textbf{03}, 033 (2011)
[arXiv:1012.4785 [hep-th]].

\bibitem{Preis:2012fh}
F.~Preis, A.~Rebhan and A.~Schmitt,
Lect. Notes Phys. \textbf{871}, 51-86 (2013)
[arXiv:1208.0536 [hep-ph]].

\bibitem{Allen:2013lda}
P.~G.~Allen and N.~N.~Scoccola,
Phys. Rev. D \textbf{88}, 094005 (2013)
[arXiv:1309.2258 [hep-ph]].

\bibitem{Allen:2015paa}
P.~G.~Allen, A.~G.~Grunfeld and N.~N.~Scoccola,
Phys. Rev. D \textbf{92}, no.7, 074041 (2015)
[arXiv:1508.04724 [hep-ph]].

\bibitem{Ferraris:2021vun}
S.~A.~Ferraris, D.~G.~Dumm, A.~G.~Grunfeld and N.~N.~Scoccola,
Eur. Phys. J. A \textbf{57}, no.4, 141 (2021)
[arXiv:2103.00982 [hep-ph]].

\bibitem{Ebert:1999ht}
D.~Ebert, K.~G.~Klimenko, M.~A.~Vdovichenko and A.~S.~Vshivtsev,
Phys. Rev. D \textbf{61}, 025005 (2000)
[arXiv:hep-ph/9905253 [hep-ph]].

\bibitem{Lu:2020rhp}
Z.~Y.~Lu, M.~L.~Du, F.~K.~Guo, U.~G.~Mei\ss{}ner and T.~Vonk,
JHEP \textbf{05} (2020), 001
doi:10.1007/JHEP05(2020)001
[arXiv:2003.01625 [hep-ph]].

\end{thebibliography}
\end{document}